\documentclass[letterpaper,11pt]{article}
\usepackage[margin=1in]{geometry}
\usepackage{amsmath,amssymb,amsthm}
\usepackage{braket}
\usepackage[ruled,vlined,linesnumbered]{algorithm2e}
\usepackage{cleveref}
\usepackage{bm}
\usepackage{mathtools}
\usepackage{enumitem}
\usepackage{booktabs}
\usepackage{graphicx}
\usepackage{subcaption} 
\usepackage{cite}
\usepackage{authblk}
\usepackage{array}
\usepackage{float}
\usepackage{longtable}

\theoremstyle{plain}
\newtheorem{remark}{Remark}

\usepackage{tikz}
\usepackage{pgfplots}
\pgfplotsset{compat=1.18} 

\begin{document}

\title{A Modular, Adaptive, and Scalable Quantum Factoring Algorithm}
	\author[1]{Alok Shukla \thanks{Corresponding author.}}
	\author[2]{Prakash Vedula}
	\affil[1]{School of Arts and Sciences, Ahmedabad University, India}
	\affil[1]{alok.shukla@ahduni.edu.in}
	\affil[2]{School of Aerospace and Mechanical Engineering, University of Oklahoma, USA}
	\affil[2]{pvedula@ou.edu}
\maketitle

\begin{abstract}
Shor's algorithm for integer factorization offers an exponential speedup over classical methods but remains impractical on Noisy Intermediate-Scale Quantum (NISQ) hardware due to the need for many coherent qubits and very deep circuits. Building on our recent work on adaptive and windowed phase-estimation methods, we have developed a modular, windowed formulation of Shor’s algorithm that potentially mitigates these limitations by restructuring phase estimation into shallow, independent circuit blocks that can be executed sequentially or in parallel, followed by lightweight classical postprocessing. This approach allows for a reduction in the size of the phase (or counting) register from a large number of qubits down to a small, fixed block size of only a few qubits (for example, three or four phase qubits were sufficient for the computational examples considered in this work), while leaving the work register requirement unchanged. 
The independence of the blocks allows for parallel execution and makes the approach more compatible with near-term hardware than the standard Shor's formulation.
An additional feature of the framework is the overlap mechanism, which introduces redundancy between blocks and enables robust reconstruction of phase information, though zero-overlap configurations can also succeed in certain regimes. 
Numerical simulations verify the correctness of the modular formulation while also showing substantial reductions in counting qubits per block.
\end{abstract}

\section{Introduction}

Shor’s algorithm for integer factorization remains one of the most striking examples of quantum advantage, offering an exponential speedup over the best-known classical factoring methods~\cite{shor1994algorithms,nielsen2002quantum}. The algorithm relies on Quantum Phase Estimation (QPE) of modular exponentiation, implemented through the quantum Fourier transform, to recover the period of a function and thereby obtain nontrivial factors of a composite integer. Beyond its foundational importance, Shor’s algorithm has profound implications for modern cryptography, since its efficient execution would compromise widely used public-key cryptographic systems.

Despite its theoretical power, the practical realization of Shor’s algorithm faces substantial barriers. In its standard form, the algorithm requires a large phase (or counting) register of $2n+1$ qubits, in addition to an $n$-qubit work register, together with long coherent circuits to implement repeated controlled modular exponentiation to factor an $n$-bit integer. Such resource demands exceed the capabilities of today’s Noisy Intermediate-Scale Quantum (NISQ) devices, which are constrained by limited qubit counts, short coherence times, and noise. A range of strategies have been explored to mitigate these requirements, including circuit-level optimizations techniques~\cite{zalka1998fast,vedral1996quantum,iqbal2024enhanced,tan2024efficient}.

Recently, we have introduced a modular and adaptive formulation of phase estimation. 
The Adaptive Windowed Quantum Phase Estimation (AWQPE) algorithm~\cite{shukla2025mqpe} decoupled estimation precision from the number of qubits in any single circuit block, enabling scalable estimation of a single eigenphase with small quantum registers. This idea was extended in the Adaptive Windowed Quantum Amplitude Estimation (AWQAE) framework~\cite{shukla2025mqae}, which demonstrated that windowing and adaptive refinement could be applied to estimation of two distinct eigenphases $\phi_1$ and $\phi_2$ (where $\phi_1 < 0.5 < \phi_2$) using a single ancilla qubit.
%

In this work, we generalize the AWQPE approach to integer factorization (that depends on quantum phase estimation involving superposition of multiple eignephases) and develop a modular version of Shor’s algorithm that is tailored to near-term architectures. A key innovation is the introduction of an overlap mechanism between blocks, together with a carry-aware stitching procedure, which allows robust reconstruction of the full phase from multiple shallow, independent circuit executions. This overlap provides redundancy that potentially improves resilience to noise and sampling fluctuations, while retaining the option to operate with zero overlap in small-block regimes.

There is also active research aimed at reducing the resource requirements of the arithmetic layer of Shor’s algorithm, including Toffoli-based modular multiplication using $2n+2$ qubits~\cite{haner2016factoring}, optimized modular exponentiation with reduced overhead~\cite{gidney2021factor}, and cryptography-inspired approaches for additional qubit savings~\cite{chevignard2025reducing}. Our work is complementary to these efforts, as it targets the phase-estimation layer, and the combination of both directions offers the potential for cumulative improvements in the overall efficiency of quantum factoring.

The remainder of the paper is organized as follows. Section~\ref{sec:methodology} describes our proposed modular approach to integer factorization based on Shor's algorithm. Section~\ref{sec:results} presents numerical demonstrations of integer factorization using shallow, quantum circuits. A deterministic candidate-based success metric is presented in \Cref{sec:deterministic-success}.
Computational complexity is discussed in Section~\ref{sec:complexity}. 
Section~\ref{sec:conclusion} concludes the paper with a summary of the results.

\section{Modular Shor's Algorithm via Windowed QPE}
\label{sec:methodology}
In this section, we present a detailed description of our proposed modular version of Shor's Algorithm, formulated via Quantum Phase Estimation (QPE) with overlapping blocks of counting qubits. This novel approach is designed to mitigate the large qubit requirements of the original Shor’s algorithm. The key idea is to partition the phase estimation procedure into a sequence of smaller, manageable quantum circuits, each responsible for a portion of the total QPE. The partial outcomes are subsequently integrated through classical post-processing to reconstruct the full phase, from which the order of \(a \bmod N\) and thereby a non-trivial factor of the composite number \(N\) can be obtained.  

Our main algorithm is summarized in \Cref{alg:main}. It requires as input a composite integer \(N\) to be factored, together with a coprime base \(a < N\). The additional parameters include the target register size \(n_{\text{target}} = \lceil \log_2 N \rceil\), a choice of block sizes \(\mathbf{m} = [m_1, \dots, m_B]\) for counting qubits, and the corresponding overlap sizes \(\mathcal{T} = [t_1, t_2, \dots, t_B]\), where \(t_1 = 0\) and, for \(i \geq 2\), each \(t_i\) satisfies \(t_i \leq \min(m_{i-1}, m_i)\).  

The phase estimation is carried out in an iterative, blockwise manner. In the first iteration, 
it is applied to the initial block of $m_1$ counting qubits, which correspond to the $m_1$ 
most significant bits (MSBs) of the phase to be estimated. In the second iteration, the procedure is repeated on a block of size $m_2$, with an enforced overlap of $t_2$ bits with the first block. Thus, the overlap region is estimated twice, i.e., once as the final $t_2$ bits of the first block, and again as the leading $t_2$ bits of the second block. As will be shown later, this redundancy 
enables consistent stitching of the measurement outcomes from independent blocks to 
reconstruct the full phase estimate. In this step, the phase estimation therefore yields 
bit positions $(m_1 - t_2 + 1)$ through $(m_1 - t_2 + m_2)$. The process continues 
analogously for all subsequent blocks. At each iteration, fresh counting qubits may be 
allocated to estimate the targeted portion of the phase, with the overall structure 
determined by the chosen block sizes 
$\mathbf{m} = [m_1, \dots, m_B]$ and overlap sizes 
$\mathcal{T} = [t_1, t_2, \dots, t_B]$.
In addition, the algorithm requires the number of measurement shots, the number of candidate outcomes to be retained, and the maximum number of candidate sequences to preserve during post-processing. Given these inputs, the algorithm outputs a non-trivial factor of \(N\).

\Cref{alg:main}  begins by partitioning the phase register into 
blocks of size $m_1, m_2, \dots, m_B$, with specified overlaps between adjacent blocks. 
In step~1, each block is processed independently: the routine 
\texttt{RunAndProcessQPEBlock} (\Cref{alg:run_block}) executes a QPE circuit for that block and returns a candidate set 
of phase bitstrings, with the exponent offset updated to align successive windows. 
In step~2, the sets from all blocks are merged by 
\texttt{CarryAwareIntegrationAndStitching} (\Cref{alg:stitch}), which uses the overlap to enforce consistency across 
block boundaries and reconstruct full-length candidate phases. 
In step~3, the routine \texttt{RecoverPeriodAndFactor} (\Cref{alg:recover_corrected}) analyzes each reconstructed candidate using 
continued fractions to extract candidate periods, verifies them against the modular exponentiation, 
and computes non-trivial factors of $N$ when possible. If a valid factor is found, the algorithm 
outputs it; otherwise, it returns that no non-trivial factor was obtained.
If a non-trivial factor is not found, the procedure can be repeated with a new random base $a$. 
In the following, we provide a more detailed description of \Cref{alg:run_block,alg:stitch,alg:recover_corrected}.

\subsection{Block-wise Quantum Phase Estimation}

\Cref{alg:run_block} sequentially (or in parallel) prepares smaller blocks of phase (or counting) qubits that interact with the modular exponentiation operator acting on the work qubits. This is in contrast with the standard QPE wherein a single large phase (or counting) register is used for phase estimation. 

In our proposed approach, a QPE circuit is constructed for each block. A key innovation is the use of an \texttt{exponent\_offset} parameter, which we denote as $\kappa_{\text{offset}}$ which is passed from \Cref{alg:main} to \Cref{alg:run_block}. We note that this parameter is incremented in each iteration in \Cref{alg:main} (in the $i$-th iteration the increment is $m_i -t_{i+1} $, ref.~line $7$ in \Cref{alg:main}). In \Cref{alg:run_block} this parameter determines the exponent of the controlled-unitary operator, $U^{2^{\kappa_j}}$, where $\kappa_j = \kappa_{\text{offset}} + j$, effectively targeting a specific window of the phase bitstring. The circuit consists of a counting register of size $m_i$ for the current block and a target register of size $n_{\text{target}}$ to hold the modular arithmetic state $|a^x \pmod{N}\rangle$. First, Hadamard gates are applied to the counting register, then for each qubit $j$, a controlled-unitary operator $U^{2^{\kappa_j}}$ is applied. This operator performs modular multiplication by $a \pmod N$. Following an inverse Quantum Fourier Transform (QFT$^{-1}$), a measurement yields a bitstring $b \in \{0,1\}^{m_i}$ that approximates the phase bits within the current window. By repeating the experiment for a number of \texttt{shots} and selecting the most frequent outcomes, a candidate set $\mathcal{S}_i$ of block-wise phase estimates is obtained. This process is repeated for each block, and the resulting candidate sets $\mathcal{S}_1, \dots, \mathcal{S}_B$ are stored for the next step.

\subsection{Carry-Aware Integration and Stitching}

Because the QPE is executed block by block, the measurement outcomes must be classically stitched together to recover a full-length estimate of the phase $\varphi=s/r$. The case we are dealing with involves a superposition of multiple eignpahses. The goal is to correctly estimate any one eigenphase, which would allow us to factor the input integer $N$ using the continued fraction algorithm. However, for each block the top candidates may come from different eigenpahses. Therefore,  a naive concatenation of block outputs would lead to errors. 

This situation can be visualized as follows. Imagine each eigenphase represented as a distinct color band (for instance, red, blue, green, or yellow). Each block outcome reveals only a short segment of these bands. Viewed in isolation, such a segment does not provide enough information to determine how it connects to the rest of the band. The overall task is to reconstruct a continuous band of the same color by combining the partial segments obtained from different blocks. A naive attempt to join individual segments may result in a multi-color band, i.e., where different segments might be of different colors. This is precisely where the notion of \emph{overlap} becomes helpful: if adjacent blocks share overlapping regions, one can verify that the colors agree within the overlap. This agreement guarantees that the stitched-together segments form a consistent, full-length band corresponding to the same eigenphase. 

Further, carry errors can propagate across the boundaries of overlapping blocks and must be accounted for while stitching the full phase string from the individual block estimates. 
To resolve this,  
\Cref{alg:stitch} provides a robust, carry-aware stitching routine. The algorithm integrates the blocks from right to left, a process that naturally accounts for the direction of carry propagation. For any two consecutive blocks, a left candidate $s_{\text{left}}$ from block $\ell$ and a right candidate $s_{\text{right}}$ from block $\ell+1$, a consistency check is performed. The check uses the overlap size, $t_{\ell+1}$, to compare the tail of the left bitstring with the head of the right bitstring, while also accounting for a potential carry bit, $c$, from the right block. This is captured by the modulo arithmetic check: $(\mathrm{int}(\mathrm{tail}(s_{\ell}, t_{\ell+1}),2) - c) \pmod{2^{t_{\ell+1}}} = \mathrm{int}(\mathrm{head}(s_{\ell+1}, t_{\ell+1}),2)$ (ref.~\Cref{remark:bitordering,remark:bitops}). If the consistency check passes, the blocks are stitched together, removing the overlap. This iterative procedure is performed for all blocks, ultimately producing a set of complete, stitched candidates, $\mathcal{C}_{\text{stitched}}$.

\subsection{Period Recovery and Factorization}

Once the full phase candidates are classically reconstructed, the final phase of the algorithm uses number-theoretic techniques to recover the period and, subsequently, the non-trivial factors of $N$.
\Cref{alg:recover_corrected} processes each stitched phase candidate. Each stitched bitstring $\hat{b}$ encodes an approximation of the phase, $\hat{y}/2^{n_{\text{total}}} \approx s/r$, where $n_{\text{total}}$ is the total number of effective phase qubits. The algorithm performs a classical continued fraction expansion of the fraction $\hat{y}/2^{n_{\text{total}}}$, which efficiently finds rational approximations. The denominators of these approximations, denoted by $\mathcal{Q}$, are strong candidates for the period $r$. Each candidate $r \in \mathcal{Q}$ is then tested to verify if it satisfies the condition $a^r \equiv 1 \pmod N$. If a valid period $r$ is found, the final classical post-processing step from Shor's algorithm is applied. This involves checking if $r$ is even and if $a^{r/2} \not\equiv -1 \pmod N$. If these conditions are met, the greatest common divisor (GCD) is computed for both $(a^{r/2}-1, N)$ and $(a^{r/2}+1, N)$. A non-trivial factor of $N$ is returned if a GCD other than 1 or $N$ is found.

\begin{remark}
\label{remark:bitordering}
\textbf{Bit-Ordering and Qiskit Convention:}
We adopt the Qiskit convention. The quantum registers are indexed LSB-first, where qubit $q[0]$ is the least significant bit (LSB) and $q[m-1]$ is the most significant bit (MSB). The controlled-unitary operator $C\text{-}U^{2^{k}}$ is applied with the control qubit $q[k]$. A measured bitstring $b$ is obtained by reading the qubits from most significant to least significant, resulting in a string
\[
b = b_{m-1} \, b_{m-2} \, \dots \, b_1 \, b_0 \quad\text{(MSB-first).}
\]
For integer conversion we treat $b$ as MSB-first and index bits by their \emph{weight}: $b_j$ denotes the bit of weight $2^j$ (so $b_{m-1}$ is the leftmost/MSB and $b_0$ is the rightmost/LSB). We define
\[
\mathrm{int}(b, 2) \;=\; \sum_{j=0}^{m-1} b_j \, 2^{\,j}.
\]
Concatenation and slicing of bitstrings are always handled as MSB-first operations.

To summarize, registers and controlled-unitary exponents use LSB indexing; measured bitstrings are MSB-first; integer conversion treats $b$ as MSB-first (no reversal); concatenation/slicing are MSB-first.
\end{remark}

\begin{remark} \label{remark_target}
In the standard Shor's algorithm, the target register is typically initialized in the state $\ket{1}$. For improved robustness and generality in our windowed QPE approach, it is advantageous to initialize the target register to a superposition of states:
\begin{equation} \label{eq_psi_general}
\ket{\psi_0} = \sum_{j=0}^{L-1} w_{j} \, \ket{a^{k_j} \bmod N}, \quad \sum_{j=0}^{L-1} |w_j|^2 = 1,
\end{equation}
where $\{k_j\}_{j=0}^{L-1}$ is a set of non-negative integer exponents chosen from a suitable range, and 
    $\{w_j\}_{j=0}^{L-1}$ are the complex weights (amplitudes) satisfying the normalization condition $\sum_{j=0}^{L-1} |w_j|^2 = 1$.

This construction yields a general multi-term superposition in the target register. The exponents \( k_j \) can be chosen in various ways, for instance, in an arithmetic progression where the exponents increase linearly (e.g., \( k_j = A + B \cdot j \)), or in a non-linear progression where they increase quadratically, exponentially, or according to any other desired non-linear pattern. The goal is to create a non-uniform initial target state such that one phase dominates in the superposition, simplifying subsequent processing.

In the special case where $L=1$ and $k_0=0$, this reduces to the conventional Shor's algorithm initialization, $\ket{\psi_0} = \ket{a^0 \bmod N} = \ket{1}$. 
Another special case is
\begin{equation} \label{eq_psi_with_one}
\ket{\psi_0} = w_0 \ket{1} + \sum_{j=1}^{L-1} w_{j} \, \ket{a^{2^{\kappa_{0}} + j -1 } \bmod N}, \quad 
\sum_{j=0}^{L-1} |w_j|^2 = 1,
\end{equation}
where $\kappa_0 = \texttt{exponent\_offset}$ (ref.~ \Cref{alg:run_block}). 

The initial state $\ket{\psi_0}$ may also be prepared as a superposition of quantum states $\ket{\widetilde{u}_j}$, which we term as the Fourier basis states. The target state is constructed as:
\begin{equation}\label{eq_psi_fourier_basis}
    \ket{\psi_0} = \sum_{j} c_j \ket{\widetilde{u}_j}, \quad \text{where } \ket{\widetilde{u}_j} = \frac{1}{\sqrt{r_j}} \sum_{s=0}^{r_j-1} e^{-2\pi i \frac{j s}{r_j}} \ket{a^s \bmod N}.
\end{equation}
For $r_j =r$, $\ket{\widetilde{u}_j}$ represents the modular exponentiation eigenstate   corresponding to the phase $\phi= j/r$.
By suitably adjusting the weight $|c_j|^2$ on specific terms, the success probability for discovering the corresponding phase $\phi= j/r$ can potentially be increased.
The values $r_j$ are trial orders or initial guesses.
\end{remark}

\begin{remark}
\label{remark:bitops}
\textbf{Bitstring Slicing Convention:}
For a bitstring $s$ of length $m$, the operation $\text{head}(s, t)$ refers to the first $t$ bits (MSB-first), while $\text{tail}(s, t)$ refers to the last $t$ bits.
\end{remark}

\begin{algorithm}[H]
\caption{Modular Shor's Algorithm via Windowed QPE}
\label{alg:main}
\SetKwInOut{Input}{Input}\SetKwInOut{Output}{Output}
\Input{A composite number $N$ to be factored; a coprime base $a < N$; 
the target register size $n_{\text{target}} = \lceil\log_2 N\rceil$; block sizes $\mathbf{m}=[m_1,\dots,m_B]$; overlap size $\mathcal{T}=[t_1,t_2\dots,t_B]$, where $t_1 =0$ and $t_i \leq \min(m_{i-1}, m_{i})$ for $i \geq 2$; number of shots $\texttt{shots}$; number of candidates to select $\texttt{num\_selected\_candidates} \geq 1$; maximum number of sequences to retain $\texttt{max\_limit\_combos}$.}
\Output{A non-trivial factor of $N$.}
\BlankLine
Initialize an empty list $\mathcal{S}$ of candidate sets\;
Initialize \texttt{exponent\_offset} $\leftarrow 0$\;
\BlankLine
\tcc{1. Run a QPE block for each block of phase qubits}
\For{$i=1$ \KwTo $B$}{
  $\mathcal{S}_i \leftarrow$ 
    \text{RunAndProcessQPEBlock}($N, a, n_{\text{target}}, m_i,$ \\
   $\qquad \qquad \qquad \qquad \qquad \qquad \qquad \texttt{exponent\_offset}$, $\texttt{shots}$, 
    $\texttt{num\_selected\_candidates})$\; 
    Append $\mathcal{S}_i$ to $\mathcal{S}$\;
  \If{$i<B$}{\texttt{exponent\_offset} $\leftarrow$ \texttt{exponent\_offset} $+\, m_i - t_{i+1}$\;}
}
 \BlankLine
\tcc{2. Integrate blocks and stitch candidates}
$\mathcal{C}_{\text{stitched}} \leftarrow \text{CarryAwareIntegrationAndStitching}(\mathcal{S}, \mathbf{m}, \mathcal{T}, \texttt{max\_limit\_combos})$\;
\BlankLine
\tcc{3. Recover period and factor N from the final candidates}
$(r, \text{factor}) \leftarrow \text{RecoverPeriodAndFactor}(\mathcal{C}_{\text{stitched}}, a, N)$\;
\If{$\text{factor}$ is found}{
    \Return{$\text{factor}$}\;
}
\BlankLine
\Return{``No non-trivial factor found"}\;
\end{algorithm}

\begin{algorithm}[H]
\caption{RunAndProcessQPEBlock}
\label{alg:run_block}
\SetKwInOut{Input}{Input}\SetKwInOut{Output}{Output}
\Input{Integer $N$, base $a$, target register size $n_{\mathrm{target}}$, block size $m_i$, shift power $\texttt{exponent\_offset}$, number of shots $\texttt{shots}$, number of candidates to select $\texttt{num\_selected\_candidates} \geq 1$.}
\Output{A candidate set $\mathcal{S}$ consisting of bit strings of size $m_i$.}
\BlankLine
\tcc{1. Construct and execute the quantum circuit}
Initialize quantum registers: a counting register $\ket{q_C} = \ket{0}^{\otimes m_i}$ and a target register $\ket{q_T} = \ket{0}^{\otimes n_{\text{target}}}$\;
Initialize the target register to the state $\ket{\psi_0}$ (mod $N$) (ref.~\cref{remark_target})\;

Apply Hadamard gates to the counting register: $\ket{q_C} \leftarrow H^{\otimes m_i} \ket{0}^{\otimes m_i}$\;
\tcc{Apply controlled unitaries for LSB-first qubits (j=0 is LSB)}
\For{$j=0$ \KwTo $m_i-1$}{
    Apply the controlled-unitary operator $U^{2^{\texttt{exponent\_offset}+j}}$ to the target register, with $\ket{q_C}[j]$ as the control qubit\;
}
Apply the inverse Quantum Fourier Transform (QFT$^{-1}$) to the counting register: $\ket{q_C} \leftarrow \mathrm{QFT}_{m_i}^{-1} \ket{q_C}$\;
Measure the counting register in the computational basis to obtain a bitstring $b$\;
\BlankLine
\tcc{2. Process measurement outcomes}
Execute the circuit for $\texttt{shots}$ total runs, recording the frequency of each outcome\;
\BlankLine
\tcc{3. Select and return candidate phases}
Select the top $\texttt{num\_selected\_candidates}$ bitstrings based on their frequencies\;
\Return the set of these bitstrings\;
\end{algorithm}

\begin{algorithm}[H]
\caption{CarryAwareIntegrationAndStitching}
\label{alg:stitch}
\SetKwInOut{Input}{Input}\SetKwInOut{Output}{Output}
\Input{A list of per-block candidate sets $\mathcal{S} = \left[\mathcal{S}_1, \dots, \mathcal{S}_B\right]$ where $\mathcal{S}_i$ contains $m_i$-bit strings (MSB-first); block sizes $\mathbf{m}=[m_1,\dots,m_B]$;  overlap size $\mathcal{T}=[t_1,t_2\dots,t_B]$, where $t_1 =0$ and $t_i \leq \min(m_{i-1}, m_{i})$ for $i \geq 2$; maximum number of sequences to retain $\texttt{max\_limit\_combos}$.}
\Output{A set of unique stitched candidates $\mathcal{C}_{\text{stitched}} = \{(\hat{b}, \hat{y}, \hat{\phi})\}$.}
\BlankLine
\tcc{1. Integrate blocks from right to left}
Initialize the set of integrated sequences $\mathcal{R} \leftarrow \{(s_B) \mid s_B \in \mathcal{S}_B\}$\;
\For{$\ell=B-1$ \KwTo $1$}{
Initialize a new set of sequences $\mathcal{R}_{\text{new}} \leftarrow \emptyset$\;
\ForEach{sequence $(s_{\text{right}}, \dots) \in \mathcal{R}$}
{\tcc{Let $s_{\text{right}}$ be the right block and $s_{\text{left}}$ the left block}
\ForEach{candidate $s_{\text{left}} \in \mathcal{S}_{\ell}$}{
\tcc{Determine the carry bit from the right block}
 Let $t = t_{\ell+1}$, and let $c$ be the bit at position $t$ of $s_{\text{right}}$ (from the  left, $0$-indexed)\;
\tcc{The condition for consistency is simplified when the overlap is zero}
\If{$t=0$ or $(\mathrm{int}(\text{tail}(s_{\text{left}}, t), 2) - c) \pmod{2^t} = \mathrm{int}(\text{head}(s_{\text{right}}, t), 2)$}{
Prepend $s_{\text{left}}$ to the current sequence: $(s_{\text{left}}, s_{\text{right}}, \dots)$\;
Add this new sequence to $\mathcal{R}_{\text{new}}$\;
\If{$|\mathcal{R}_{\text{new}}| \geq \texttt{max\_limit\_combos}$}{break\;}
}
}
\If{$|\mathcal{R}_{\text{new}}| \geq \texttt{max\_limit\_combos}$}{break\;}
}
$\mathcal{R} \leftarrow \mathcal{R}_{\text{new}}$\;
 \If{$\mathcal{R}$ is empty}{break\;}
}
\BlankLine
\tcc{2. Stitch the final sequences}
Initialize a set of unique stitched candidates $\mathcal{C}_{\text{stitched}} \leftarrow \emptyset$\;
\ForEach{integrated sequence $(s_1, \dots, s_B) \in \mathcal{R}$}{
 Compute $n_{\text{total}} \leftarrow \sum_{i=1}^{B} (m_i - t_i)$\;
\tcc{Concatenate the non-overlapping heads of each block from left to right.}
 Concatenate with overlap removal: $\hat{b} \leftarrow \text{head}(s_1, m_1 - t_1) \vert \text{head}(s_{2}, m_2 - t_2) \vert \cdots \vert \text{head}(s_{B}, m_B - t_B)$\;
\tcc{`int' converts a bitstring to its integer representation based on the MSB-first convention.}
 Convert to integer and phase: $\hat{y} \leftarrow \mathrm{int}(\hat{b}, 2)$, and $\hat{\phi} \leftarrow \hat{y} / 2^{\,n_{\text{total}}}$\;
Add the unique tuple $(\hat{b},\hat{y},\hat{\phi})$ to $\mathcal{C}_{\text{stitched}}$\;
}
\BlankLine
\Return{$\mathcal{C}_{\text{stitched}}$}\;
\end{algorithm}

\begin{algorithm}[H]
\caption{RecoverPeriodAndFactor}
\label{alg:recover_corrected}
\SetKwInOut{Input}{Input}\SetKwInOut{Output}{Output}
\Input{Stitched candidates $\mathcal{C}_{\text{stitched}}$; base $a$ coprime to $N$.}
\Output{A tuple $(r, \text{factor})$ where $\text{factor}$ is a non-trivial factor of $N$, or $(\texttt{None}, \texttt{None})$ if no factor is found.}
\BlankLine
\tcc{Per-Candidate Period Estimates via Continued Fractions}
\ForEach{$(\hat{b}, \hat{y}, \hat{\phi}) \in \mathcal{C}_{\text{stitched}}$}{
    Let $n_{\text{current\_total}} \leftarrow |\hat{b}|$\;
   Compute denominators $\mathcal{Q}$ from the convergents of 
$\hat{y} / 2^{n_{\text{current\_total}}}$ via continued fractions, with $q \leq N$.
\;
    \ForEach{$r \in \mathcal{Q}$}{
        \If{$r \leq 0$ or $r > N$}{
            continue\;
        }
        \tcc{Directly check if this is the period}
        \If{$a^{r} \equiv 1 \pmod N$}{
            \tcc{Classical Post-Processing for Factors}
            \If{$r$ is even and $a^{r/2} \not\equiv -1 \pmod N$}{
                $f_1 \leftarrow \text{gcd}(a^{r/2}-1, N)$\;
                $f_2 \leftarrow \text{gcd}(a^{r/2}+1, N)$\;
                \If{$f_1 \neq 1$ and $f_1 \neq N$}{
                    \Return{$(r, f_1)$}\;
                }
                \If{$f_2 \neq 1$ and $f_2 \neq N$}{
                    \Return{$(r, f_2)$}\;
                }
            }
        }
    }
}
\BlankLine
\Return $(\texttt{None}, \texttt{None})$\;
\end{algorithm}

\subsection{Rationale for Using a Non-Standard Initial Target State}
\label{sec:nonstandard_target}

In the standard formulation of Shor’s algorithm, the target register is initialized in the computational basis state $\ket{1}$. 
This initialization leads to a highly uniform probability distribution in the measured phase register.

To enhance robustness and spectral diversity in our adaptive and windowed QPE framework, as discussed in \Cref{remark_target}, we prepare a more general superposition in the target register:
\begin{equation*} 
\ket{\psi_0} = \sum_{j=0}^{L-1} w_{j} \, \ket{a^{k_j} \bmod N}, 
\qquad 
\sum_{j=0}^{L-1} |w_j|^2 = 1,
\end{equation*}
where $\{k_j\}_{j=0}^{L-1}$ is a set of appropriately chosen non-negative integer exponents. The exponents $\{k_j\}$ and amplitudes $\{w_j\}$ can be selected to produce a non-uniform distribution over eigenstates, resulting in one or more dominant eigencomponents. This, in turn, can improve the distinguishability of valid phase estimates.

\vspace{0.4em}
\noindent\textbf{Connection to the eigenbasis.}
Let $\{\ket{u_k}\}_{k=0}^{r-1}$ denote the eigenstates of $U_a$, satisfying
$U_a\ket{u_k} = e^{2\pi i k / r}\ket{u_k}$, where  $\ket{u_k} = \frac{1}{\sqrt{r}} \sum_{s=0}^{r-1} e^{-2\pi i k s / r} \ket{a^s \bmod N}$.
For any state $ \ket{\chi}= \sum_{s=0}^{r-1} w_s \ket{a^s \bmod N}$, one can also express it in terms of eigenstates $\ket{u_k}$ as $\ket{\chi} =  \sum_{k=0}^{r-1} c_k \ket{u_k}$, where the eigen-basis amplitudes are given by
$c_k = \langle u_k | \chi \rangle = \frac{1}{\sqrt{r}} \sum_{s=0}^{r-1} w_s e^{2\pi i k s / r}$.
In particular, $\ket{\psi_0}$ also has a  spectral decomosition in terms of  eigenstates $\ket{u_k}$.
Importantly, the modified initialization does not alter the underlying eigenphases $2\pi k / r$, 
but only alters their relative probabilities in the corresponding measurement outcomes.

\vspace{0.4em}
\noindent\textbf{Example ($L=2$).}
For a simple two-term initialization
\[
\ket{\psi_0} = \frac{1}{\sqrt{2}}\big(\ket{1} + \ket{a} \big),
\]
we have $w_0 = w_1 = 1/\sqrt{2}$ and $w_{s\ge2}=0$.
Using the above formulation, the decomposition becomes
\[
\ket{\psi_0} = \sum_{k=0}^{r-1} 
  \frac{1}{\sqrt{2r}}\big(1 + e^{2\pi i k / r}\big)\ket{u_k},
\]
yielding spectral weights
\[
|c_k|^2 = \frac{2}{r}\cos^2\!\left(\frac{\pi k}{r}\right).
\]
Thus, the modified initialization redistributes amplitude non-uniformly across the eigenstates of $U_a$,
producing a distinct, structured pattern in the observed phase histogram.
This structured modulation remains robust (across blocks) and enhances correlation between blocks (with overlap) and supports partial-block stitching under limited resolution,
making it particularly advantageous in our windowed QPE approach.

\vspace{0.5em}

\begin{remark}
For simplicity \texttt{num\_selected\_candidates} is taken to be identical for all blocks, though in general each block may use its own value, for example \(N_i^{\mathrm{sel}}\) for the \(i\)-th block.
\end{remark}

\begin{remark}
\label{remark_lcm}
Further improvements to robustness in the overall order-finding process can be achieved by combining multiple phase candidates arising from different initialization choices or independent runs using the least common multiple (LCM) of their corresponding period estimates.
Since the initialization in \Cref{eq_psi_with_one} affects only the probability distribution and not the eigenphases themselves, 
the valid period $r$ remains consistent across such runs.
Taking the LCM of independently obtained candidate periods amplifies stability and helps suppress spurious estimates introduced by measurement noise or imperfect block stitching.
\end{remark}

\section{Numerical Examples}
\label{sec:results}

To illustrate our proposed modular approach to Shor’s algorithm (\Cref{alg:main}), we present several 
numerical examples. These small-scale demonstrations verify the correctness of the approach 
and highlight how varying block sizes and overlaps affect the phase-estimation procedure, 
resource requirements, and the subsequent recovery of non-trivial factors. 

For all the examples considered in this section we assume $L=1$ and $\ket{\psi_0} = \ket{1}$, unless otherwise specified (ref.~\Cref{remark_target}). 
We begin with the 
case $N=15$, and then consider alternative block configurations 
and a larger composite, $N=221$.

\subsection{Numerical Example: Quantum Factoring of \(N=15\)}

 We select a random base \(a=2\), for which the modular order for $N=15$ is known to be $r=4$. A standard QPE circuit requires at least $2\lceil\log_2 N\rceil+1=9$ qubits to guarantee a correct result. Here, we show a successful run using a windowed QPE approach with overlapping blocks and a reduced total effective phase-register length per block. For this run, the parameters for the stitching procedure were set as follows: \texttt{max\_limit\_combos} = 2 and \texttt{num\_selected\_candidates} = 2.

\noindent \paragraph{Quantum Phase Estimation with Overlapping Blocks.}

The QPE is executed in four sequential blocks, denoted as \(B=4\). The block sizes are defined by the vector $\mathbf{m}= [m_1,m_2,m_3,m_4] = [3, 4, 4, 5]$, while the corresponding overlaps are given by $\mathcal{T}= [t_1, t_2, t_3, t_4] = [0, 2, 3, 2]$. For each block, a circuit is run to measure the most significant qubits of the phase register. Table \ref{tab:block_outcomes_15} summarizes the most frequent outcomes, with probabilities derived empirically from simulation shots.

\begin{table}[ht]
\centering
\caption{Most frequent candidates measured during the run for $N=15$ with $a=2$.}
\label{tab:block_outcomes_15}
\begin{tabular}{c c p{9cm}}
\toprule
{Block} & {Block Size} ($m_i$) & {Top Candidates (bitstring: empirical probability)} \\
\midrule
1 & $3$ & \texttt{110} (29.0\%), \texttt{010} (27.0\%) \\
2 & $4$ & \texttt{1000} (51.0\%), \texttt{0000} (49.0\%) \\
3 & $4$ & \texttt{0000} (100.0\%) \\
4 & $5$ & \texttt{00000} (100.0\%) \\
\bottomrule
\end{tabular}
\end{table}

\begin{figure*}[ht!]
    \centering
    \begin{subfigure}[b]{0.45\textwidth}
        \centering
        \includegraphics[width=\textwidth]{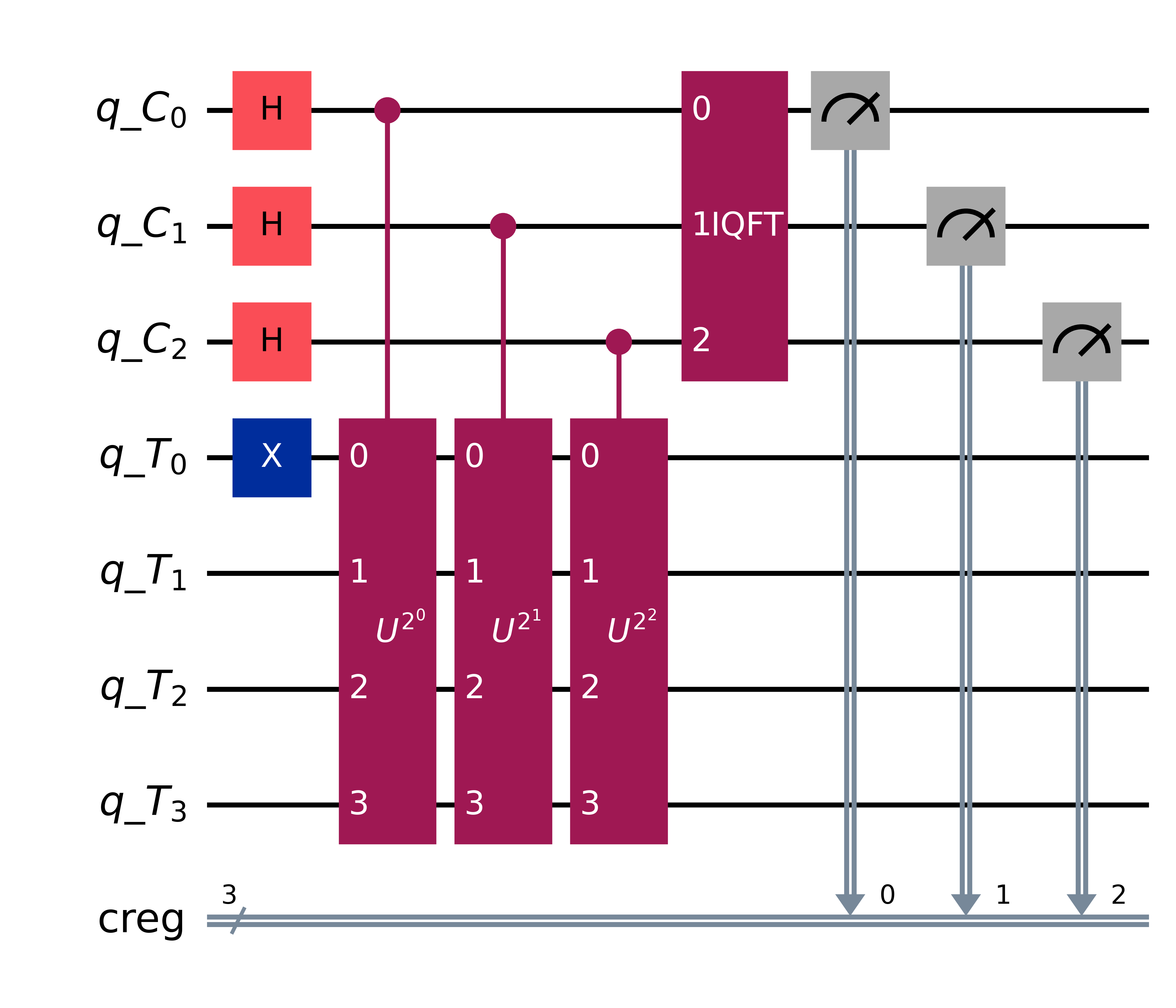}
        \caption{Quantum circuit for Block 1 with block size $m_1=3$.}
        \label{fig:block1_circuit}
    \end{subfigure}
    \hfill
    \begin{subfigure}[b]{0.45\textwidth}
        \centering
        \includegraphics[width=\textwidth]{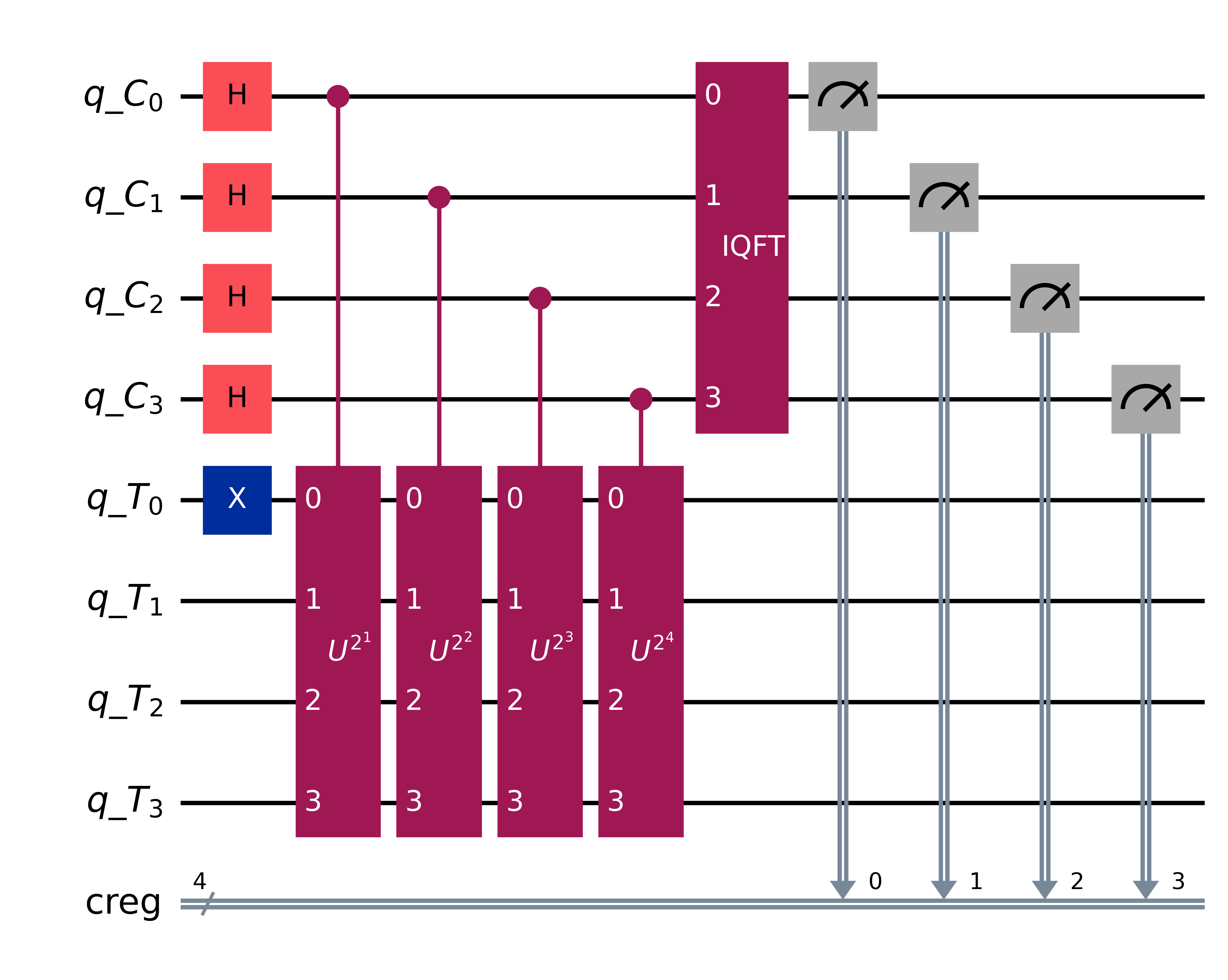}
        \caption{Quantum circuit for Block 2 with block size $m_2=4$.}
        \label{fig:block2_circuit}
    \end{subfigure}
    \par\bigskip 
    \begin{subfigure}[b]{0.45\textwidth}
        \centering
        \includegraphics[width=\textwidth]{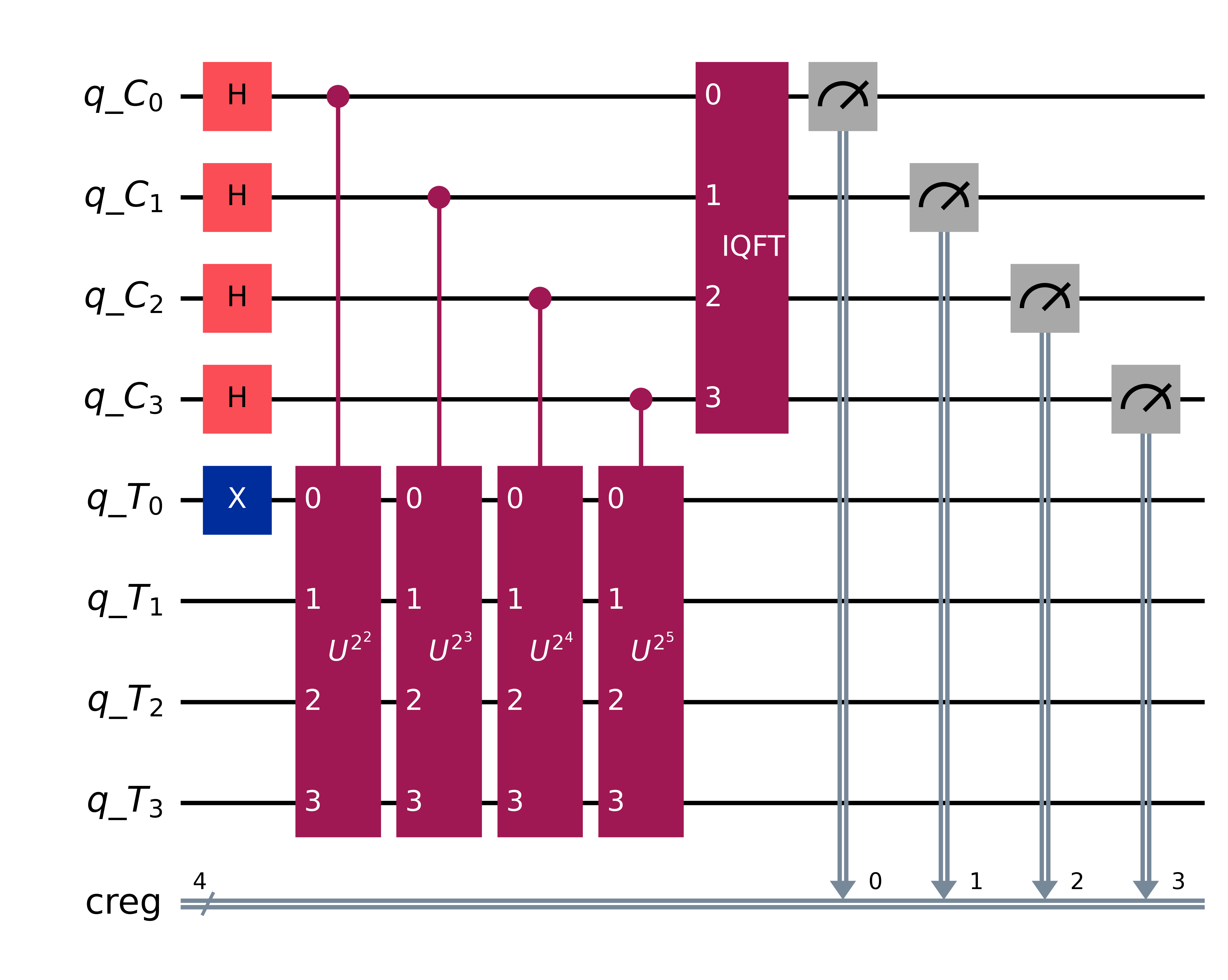}
        \caption{Quantum circuit for Block 3 with block size $m_3=4$.}
        \label{fig:block3_circuit}
    \end{subfigure}
    \hfill
    \begin{subfigure}[b]{0.45\textwidth}
        \centering
        \includegraphics[width=\textwidth]{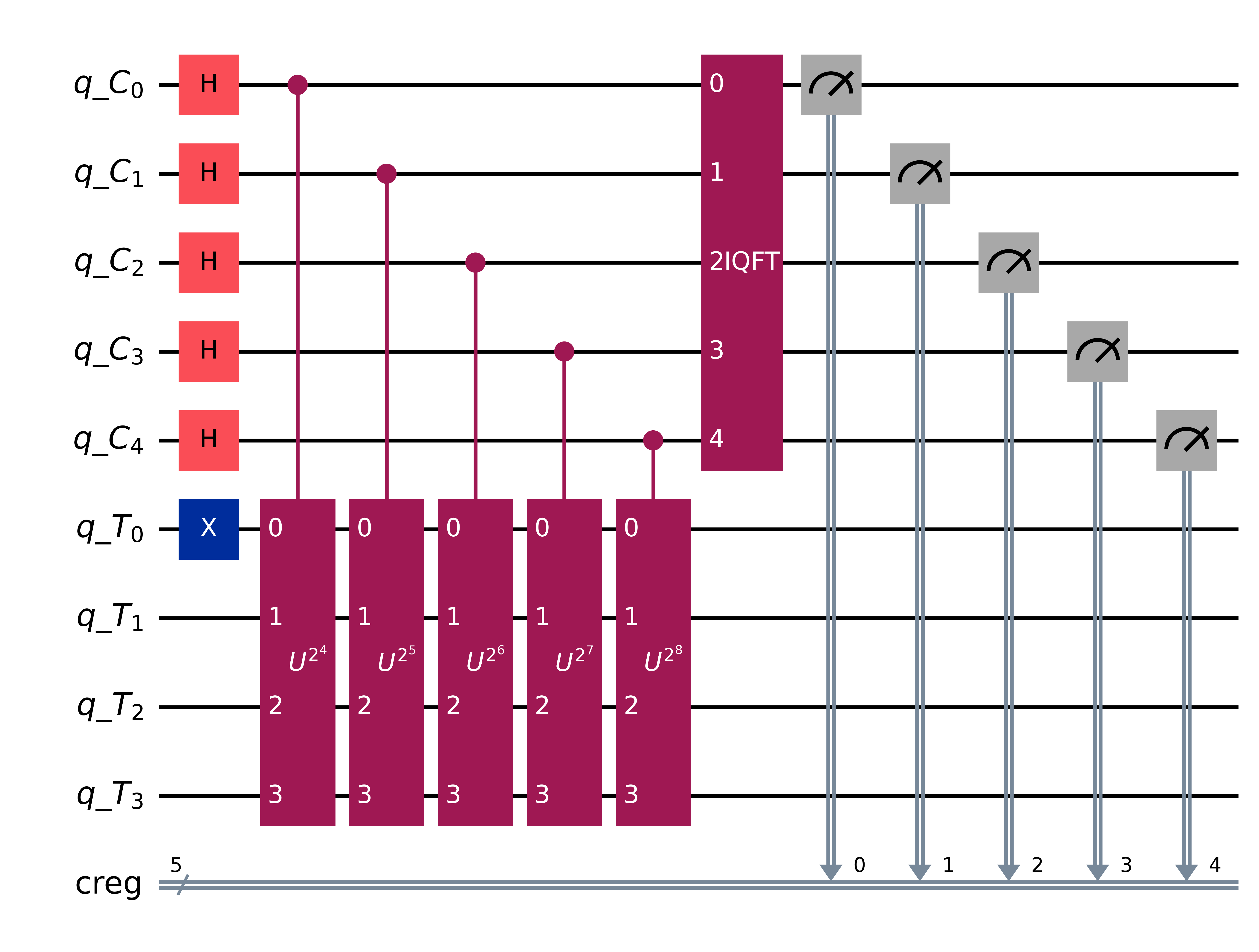}
        \caption{Quantum circuit for Block 4 with block size $m_4=5$.}
        \label{fig:block4_circuit}
    \end{subfigure}
    \caption{Quantum circuits for each of the four blocks in the windowed QPE run for factoring $N=15$.}
    \label{fig:all_circuits}
\end{figure*}

\noindent \paragraph{Quantum Circuit Diagrams for Each Block.}
\Cref{fig:all_circuits} shows the quantum circuit for each block of the windowed QPE run (ref.~\Cref{alg:main,alg:run_block}).

\noindent \paragraph{Carry-Aware Integration and Stitching.}

Following the quantum measurements, a classical stitching procedure reconstructs the full-length phase bitstring (ref.~\Cref{alg:stitch}). This process operates from right-to-left, beginning with the least significant bits (LSBs) measured in the final QPE block and progressing towards the most significant bits (MSBs).
The total length of the stitched bitstring is 9 bits, a sum of the non-overlapping segments from each block:
\begin{equation*}
 (m_1-t_1) + (m_2-t_2) + (m_3-t_3) + (m_4-t_4) = (3-0) + (4-2) + (4-3) + (5-2) = 3+2+1+3 = 9.
\end{equation*}
Using the most probable outcome from each block, we obtain the stitched bitstring \texttt{110000000}. The process initiates with the non-overlapping portion of Block 4's outcome, taking the final $m_4-t_4=3$ bits (\texttt{000}). Next, the non-overlapping portion of Block 3's outcome, the final $m_3-t_3=1$ bit, is prepended (\texttt{0000}). We then prepend the non-overlapping portion of Block 2's outcome, the final $m_2-t_2=2$ bits (\texttt{000000}). Finally, the non-overlapping portion of Block 1's outcome, the final $m_1-t_1=3$ bits, is prepended to complete the bitstring (\texttt{110000000}).
This concatenation yields the full stitched bitstring, which corresponds to two unique phase candidates. The first, derived from the second most probable outcome of the initial block, is the bitstring \texttt{010000000}, which represents the integer $\hat{y} = \texttt{010000000}_2 = 128$ and a phase of $\hat{\phi} = \frac{\hat{y}}{2^9} = \frac{1}{4}$. The second candidate, derived from the most probable outcomes, is the bitstring \texttt{110000000}, which corresponds to the integer $\hat{y} = \texttt{110000000}_2 = 384$ and a phase of $\hat{\phi} = \frac{\hat{y}}{2^9} = \frac{3}{4}$.

\noindent \paragraph{Period Recovery and Factorization.}

The final stage employs the classical continued fraction algorithm on the observed phase, $\hat{\phi}=3/4$~ (ref.~\Cref{alg:recover_corrected}). This yields the most likely denominator $r=4$, which is confirmed to be the true period of 2 modulo 15. The factorization is then completed by computing the greatest common divisor (gcd):
\begin{equation}
\label{eq:factor_15}
\text{gcd}\left(a^{r/2} \pm 1, N\right) = \text{gcd}\left(2^{4/2} \pm 1, 15\right) = \text{gcd}\left(2^{2} \pm 1, 15\right) \Rightarrow \{3, 5\}. \nonumber
\end{equation}
The algorithm successfully returns the non-trivial factors 3 and 5.

\subsection{Numerical Example: Quantum Factoring of \(N=15\) with Fewer Qubits}

A standard QPE circuit requires at least $2\lceil\log_2 N\rceil+1=9$ qubits to guarantee a correct result. However, in some special cases, successful factorization can be achieved using fewer number of qubits.  Here, we show a successful run using a windowed QPE approach with a reduced total effective phase-register length of 5 qubits, achieved using only two measurement blocks. For this run, the following parameters were used: $a = 2$, \texttt{max\_limit\_combos} = 2 and \texttt{num\_selected\_candidates} = 2.

\noindent \paragraph{Quantum Phase Estimation with Overlapping Blocks.}

The QPE is executed in two independent blocks, denoted as \(B=2\). The block sizes are defined by the vector $\mathbf{m}=[3, 4]$, while the corresponding overlaps are given by $\mathcal{T}=[0, 2]$. For each block, a circuit is run to measure the most significant qubits of the phase register. Table \ref{tab:block_outcomes_15_fewer} summarizes the most frequent outcomes, with probabilities derived empirically from simulation shots.

\begin{table}[ht]
\centering
\caption{Most frequent candidates measured during the run for $N=15$ with $a=2$.}
\label{tab:block_outcomes_15_fewer}
\begin{tabular}{c c p{9cm}}
\toprule
{Block} & {Block Size} ($m_i$) & {Top Candidates (bitstring: empirical probability)} \\
\midrule
1 & $3$ & \texttt{100} (27.0\%), \texttt{110} (25.0\%) \\
2 & $4$ & \texttt{1000} (51.0\%), \texttt{0000} (49.0\%) \\
\bottomrule
\end{tabular}
\end{table}

\noindent \paragraph{Carry-Aware Integration and Stitching.}

In this case, the total length of the stitched bitstring is 5 bits, a sum of the non-overlapping segments from each block:
\begin{equation*}
\text{Length} = (m_1-t_1) + (m_2-t_2) = (3-0) + (4-2) = 3+2 = 5.
\end{equation*}
Using the most probable outcome from each block, we reconstruct the stitched bitstring. The process initiates with the non-overlapping portion of Block 2's outcome, taking the final $m_2-t_2=2$ bits (\texttt{00}). We then prepend the non-overlapping portion of Block 1's outcome, the final $m_1-t_1=3$ bits, to complete the bitstring (\texttt{11000}).
This concatenation yields the full stitched bitstring, which corresponds to two unique phase candidates. The first, derived from the second most probable outcome of the initial block, is the bitstring \texttt{10000}, which corresponds to the integer 16 and a phase of $\frac{1}{2}$. The second candidate, derived from the most probable outcomes, is the bitstring \texttt{11000}, which represents the integer 24 and a phase of $\frac{3}{4}$.

\noindent \paragraph{Period Recovery and Factorization.}

The final stage uses the classical continued fraction algorithm on the observed phase, $\hat{\phi}=3/4$. This gives the most likely denominator $r=4$, which is confirmed to be the true period of 2 modulo 15. The factorization is then completed by computing the greatest common divisor (gcd):
\begin{equation}
\label{eq:factor_15_fewer}
\text{gcd}\left(a^{r/2} \pm 1, N\right) = \text{gcd}\left(2^{4/2} \pm 1, 15\right) = \text{gcd}\left(2^{2} \pm 1, 15\right) \Rightarrow \{3, 5\}. \nonumber
\end{equation}
The algorithm successfully returns the non-trivial factors 3 and 5.

\subsection{Numerical Example: Quantum Factoring of \(N=15\) with Zero Overlap}

Here, we show an implementation using a windowed QPE approach with zero-overlap and a reduced total effective phase-register length of 6 qubits. We select a random base $a=2$ for this demonstration. Further, the parameters for the stitching procedure were set as follows: \texttt{max\_limit\_combos} = 4 and \texttt{num\_selected\_candidates} = 4. It is important to note that the overlap vector is configured as all zeros, which disables the algorithm's built-in error mitigation and redundancy checks.

This example, with a small number of blocks, shows that a zero-overlap configuration can still produce the correct result. However, in cases with many blocks, a large number of candidates would be produced, leading to significant classical overhead. The total number of combinations to check grows exponentially with the number of blocks. This approach may still be preferable if the classical computing requirements are manageable for a given problem instance.

\noindent \paragraph{Quantum Phase Estimation with Non-overlapping Blocks.}

The QPE is executed in two sequential blocks, denoted as \(B=2\). The block sizes are defined by the vector $\mathbf{m}=[3, 3]$, while the corresponding overlaps are given by $\mathcal{T}=[0, 0]$ (i.e. the blocks do not overlap). For each block, a circuit is run to measure the most significant qubits of the phase register. Table \ref{tab:block_outcomes_15_no_overlap_2} summarizes the most frequent outcomes, with probabilities derived empirically from simulation shots.

\begin{table}[ht]
\centering
\caption{Most frequent candidates measured during the run for $N=15$ with $a=2$.}
\label{tab:block_outcomes_15_no_overlap_2}
\begin{tabular}{c c p{9cm}}
\toprule
{Block} & {Block Size} ($m_i$) & {Top Candidates (bitstring: empirical probability)} \\
\midrule
1 & $3$ & \texttt{100} (29.0\%), \texttt{110} (29.0\%), \texttt{000} (24.0\%), \texttt{010} (18.0\%) \\
2 & $3$ & \texttt{000} (100.0\%) \\
\bottomrule
\end{tabular}
\end{table}

\noindent \paragraph{Concatenation and Phase Estimation.}

Given the lack of overlap, the classical stitching procedure simplifies to a direct concatenation of the most probable bitstrings from each block. This process does not allow for error correction between blocks.
The total length of the stitched bitstring is 6 bits, a direct sum of the block sizes:
\begin{equation*}
\text{Length} = m_1 + m_2 = 3 + 3 = 6.
\end{equation*}
This concatenation yields four unique phase candidates. The most probable outcome, derived from the most probable bitstrings of each block, is the bitstring \texttt{100000}, which corresponds to the integer 32 and a phase of $\frac{1}{2}$. Other stitched candidates include \texttt{000000} (integer 0, phase 0), \texttt{010000} (integer 16, phase $\frac{1}{4}$), and \texttt{110000} (integer 48, phase $\frac{3}{4}$). The measured phase that yields the correct period is $\hat{\phi} = \frac{3}{4}$, which corresponds to the stitched bitstring \texttt{110000} from the second-most probable outcome of Block 1.

\noindent \paragraph{Period Recovery and Factorization.}

The final stage employs the classical continued fraction algorithm on the observed phase, $\hat{\phi}=3/4$. This gives the most likely denominator $r=4$, which is confirmed to be the true period of 2 modulo 15. The factorization is then completed by computing the greatest common divisor (gcd):
\begin{equation}
\label{eq:factor_15_no_overlap_2}
\text{gcd}\left(a^{r/2} \pm 1, N\right) = \text{gcd}\left(2^{4/2} \pm 1, 15\right) = \text{gcd}\left(2^{2} \pm 1, 15\right) \Rightarrow \{3, 5\}. \nonumber
\end{equation}
The algorithm successfully returns the non-trivial factors 3 and 5.

\subsection{Numerical Example: Quantum Factoring of \(N=221\)}

Next we illustrate our approach by considering the example of factoring  \(N=221\). We select a random base \(a=12\), for which the modular order is known to be $r=16$. A standard QPE circuit requires at least $2\lceil\log_2 N\rceil+1=17$ qubits to guarantee a correct result. Here, we show a successful run using a windowed QPE approach with a reduced total effective phase-register length of 7 qubits. For this run, the parameters for the stitching procedure were set as follows: \texttt{max\_limit\_combos} = 4 and \texttt{num\_selected\_candidates} = 4.

\noindent \paragraph{Quantum Phase Estimation with Overlapping Blocks.}

The QPE is executed in four sequential blocks, denoted as \(B=4\). The block sizes are defined by the vector $\mathbf{m}=[3, 3, 4, 3]$, while the corresponding overlaps are given by $\mathcal{T}=[0, 2, 2, 2]$. For each block, a circuit is run to measure the most significant qubits of the phase register. Table \ref{tab:block_outcomes_221} summarizes the most frequent outcomes, with probabilities derived empirically from simulation shots.

\begin{table}[ht]
\centering
\caption{Most frequent candidates measured during the run for $N=221$ with $a=12$.}
\label{tab:block_outcomes_221}
\begin{tabular}{c c p{9cm}}
\toprule
{Block} & {Block Size} ($m_i$) & {Top Candidates (bitstring: empirical probability)} \\
\midrule
1 & $3$ & \texttt{010} (18.0\%), \texttt{000} (18.0\%), \texttt{111} (16.0\%), \texttt{001} (14.0\%) \\
2 & $3$ & \texttt{011} (18.0\%), \texttt{001} (17.0\%), \texttt{000} (16.0\%), \texttt{111} (16.0\%) \\
3 & $4$ & \texttt{0100} (33.0\%), \texttt{1100} (26.0\%), \texttt{0000} (24.0\%), \texttt{1000} (17.0\%) \\
4 & $3$ & \texttt{000} (100.0\%) \\
\bottomrule
\end{tabular}
\end{table}

\begin{figure*}
    \centering
    \begin{subfigure}[b]{0.45\textwidth}
        \centering
        \includegraphics[width=\textwidth]{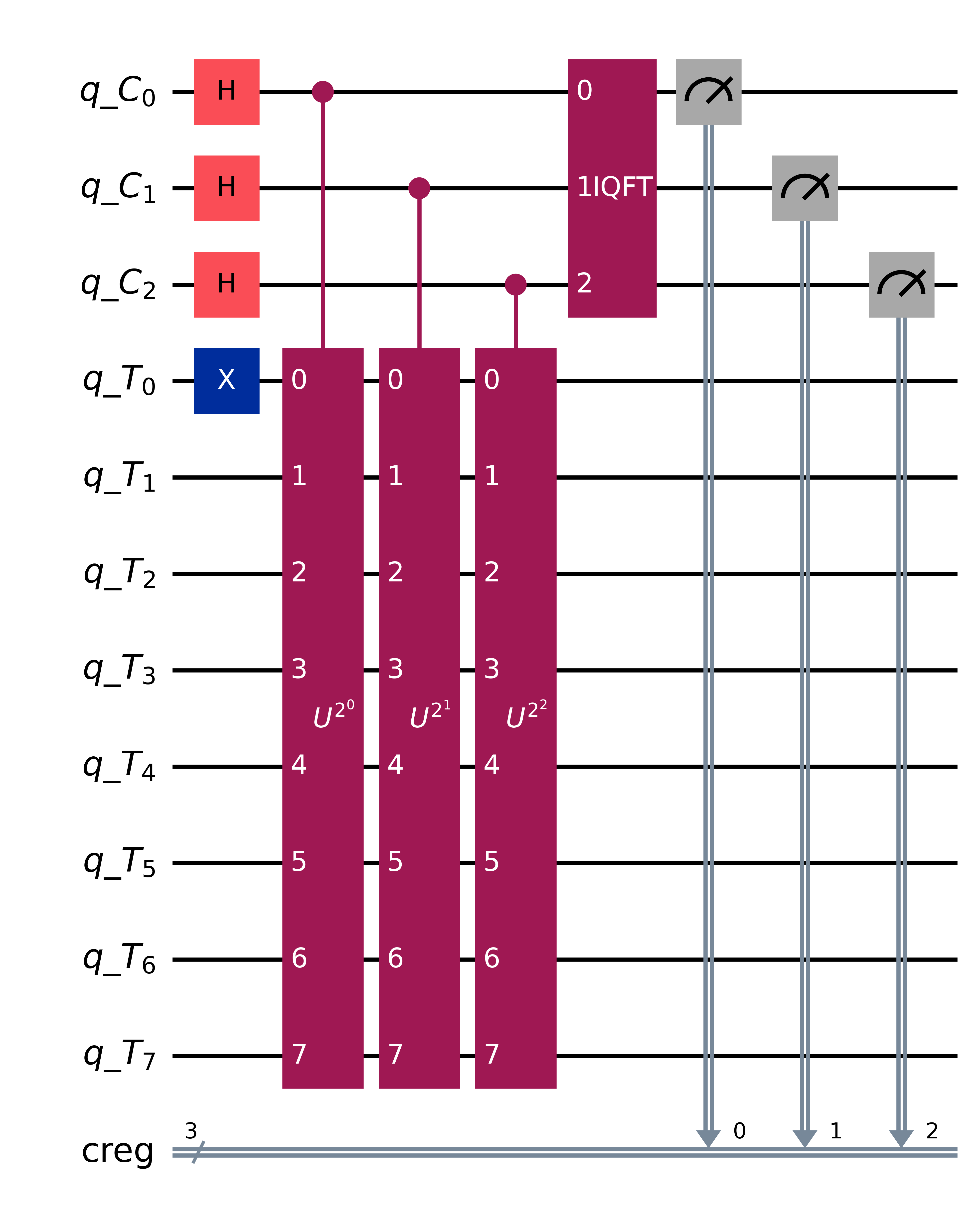}
        \caption{Quantum circuit for Block 1 with block size $m_1=3$.}
        \label{fig:block1_circuit_221}
    \end{subfigure}
    \hfill
    \begin{subfigure}[b]{0.45\textwidth}
        \centering
        \includegraphics[width=\textwidth]{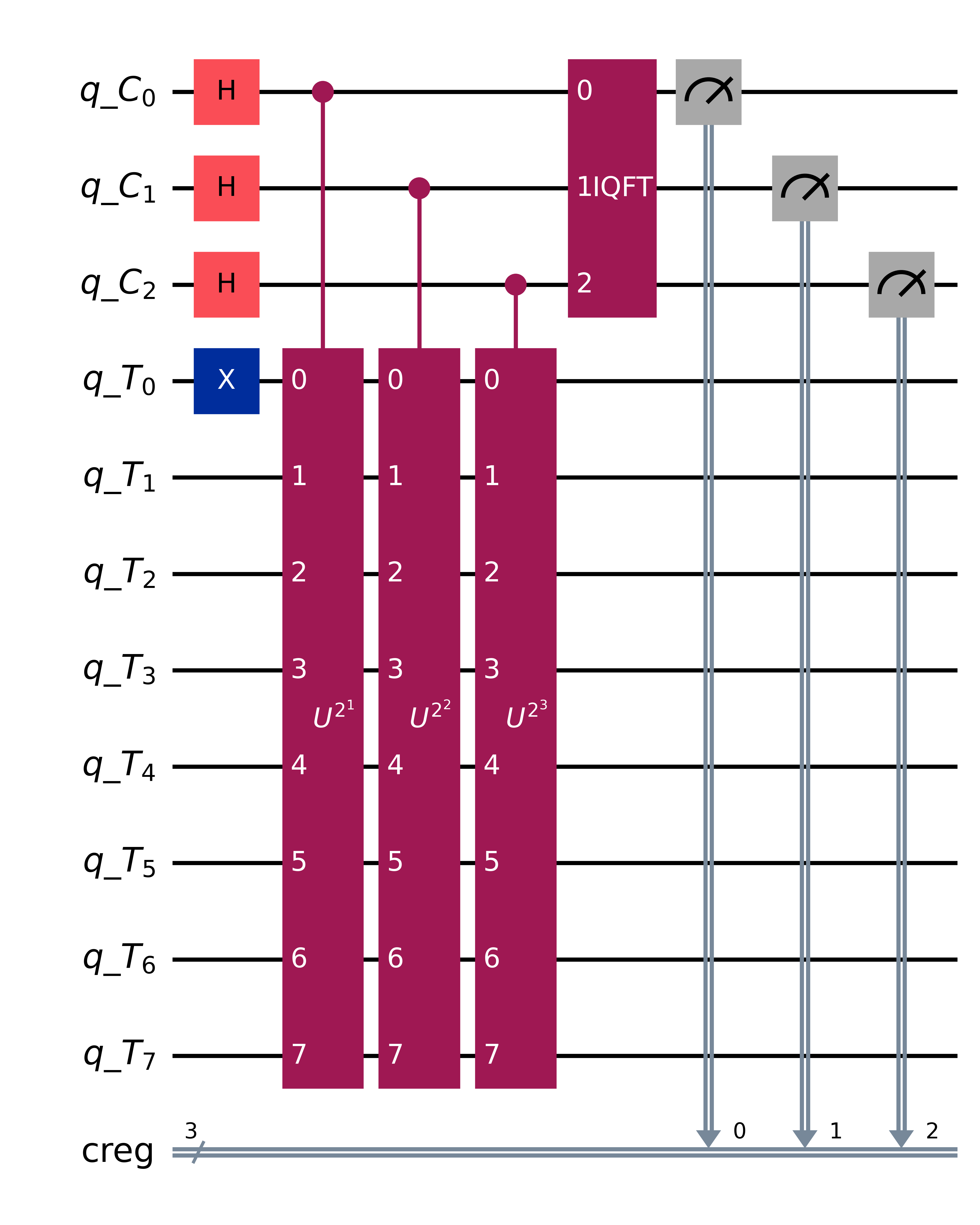}
        \caption{Quantum circuit for Block 2 with block size $m_2=3$.}
        \label{fig:block2_circuit_221}
    \end{subfigure}
    \par\bigskip 
    \begin{subfigure}[b]{0.45\textwidth}
        \centering
        \includegraphics[width=\textwidth]{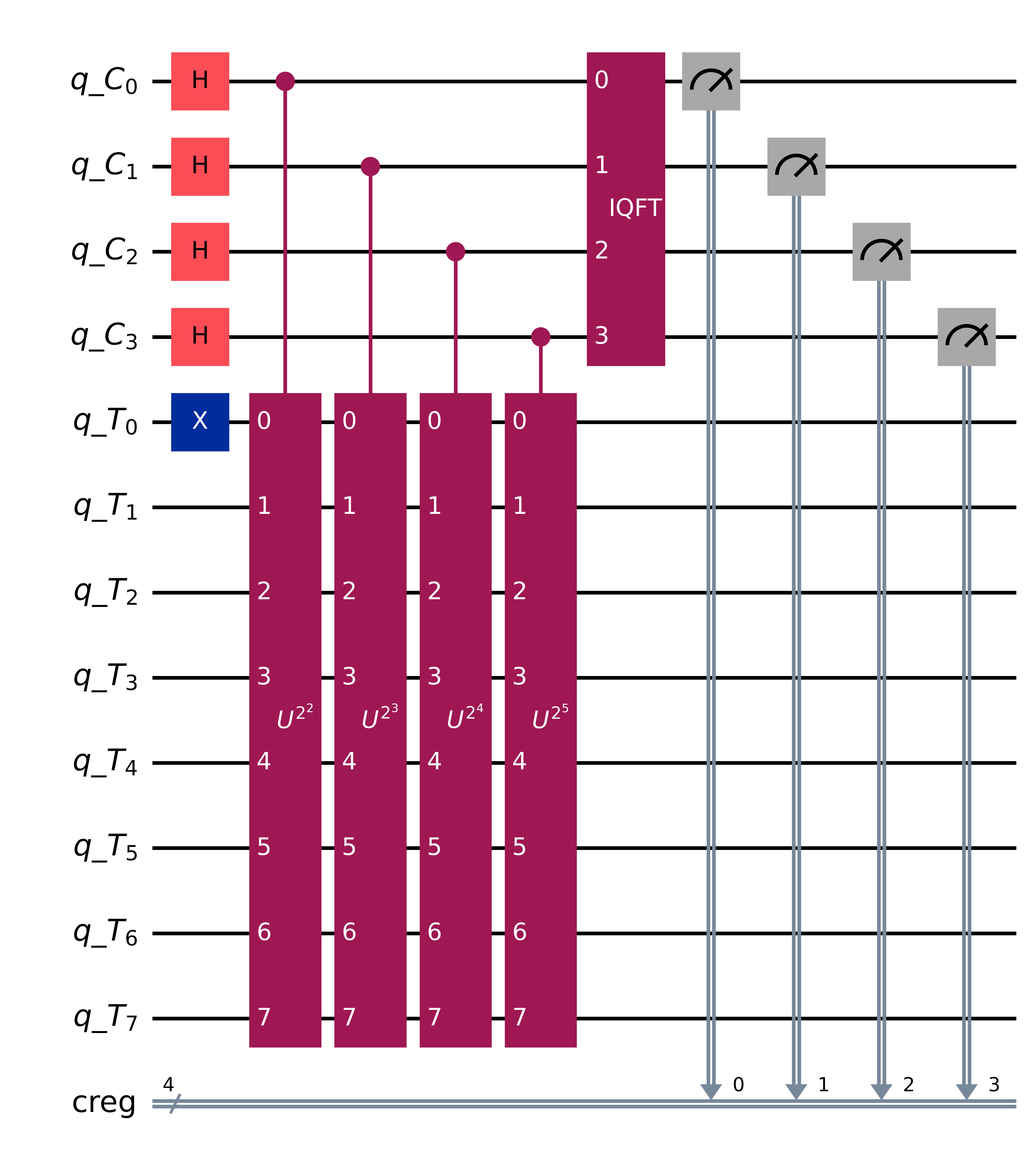}
        \caption{Quantum circuit for Block 3 with block size $m_3=4$.}
        \label{fig:block3_circuit_221}
    \end{subfigure}
    \hfill
    \begin{subfigure}[b]{0.45\textwidth}
        \centering
        \includegraphics[width=\textwidth]{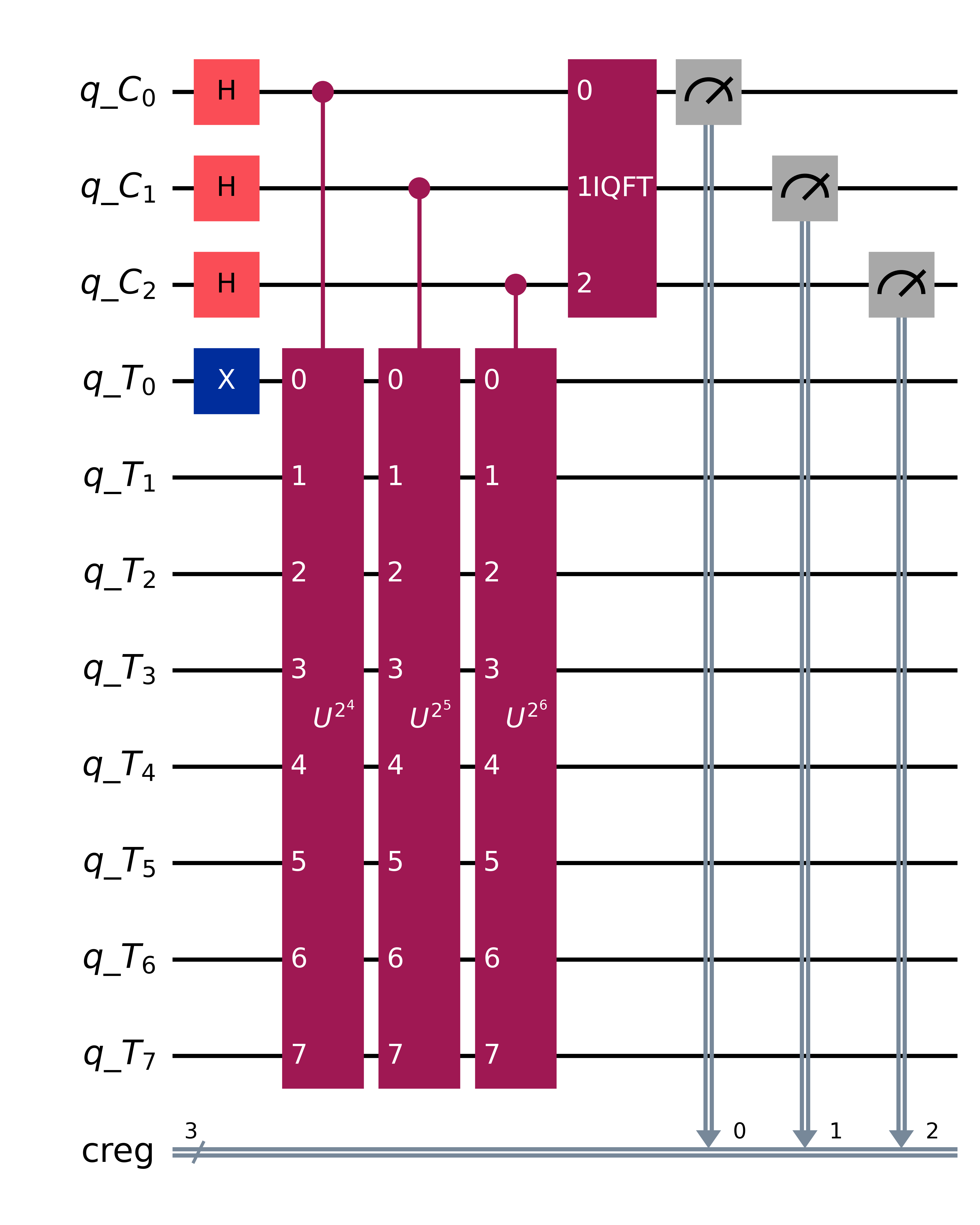}
        \caption{Quantum circuit for Block 4 with block size $m_4=3$.}
        \label{fig:block4_circuit_221}
    \end{subfigure}
    \caption{Quantum circuits for each of the four blocks in the windowed QPE run for factoring $N=221$.}
    \label{fig:all_circuits_221}
\end{figure*}

\noindent \paragraph{Quantum Circuit Diagrams for Each Block.}
\Cref{fig:all_circuits_221} shows the quantum circuit for each block of the windowed QPE run for $N=221$. (ref.~\Cref{alg:main,alg:run_block}).

\noindent \paragraph{Carry-Aware Integration and Stitching.}

In this case, the total length of the stitched bitstring is 7 bits, a sum of the non-overlapping segments from each block:
\begin{equation*}
\text{Length} = (m_1-t_1) + (m_2-t_2) + (m_3-t_3) + (m_4-t_4) = (3-0) + (3-2) + (4-2) + (3-2) = 3+1+2+1 = 7.
\end{equation*}

Four unique phase candidates were obtained through the carry-aware integration and stitching algorithm using the most probable outcomes from each block.
The optimal candidate derived from this process is the bitstring \texttt{0111000}, which corresponds to the integer 56 and a phase of $\frac{7}{16}$. 
Other stitched candidates include \texttt{0000000} (integer 0, phase 0), \texttt{0001000} (integer 8, phase $\frac{1}{16}$), and \texttt{0011000} (integer 24, phase $\frac{3}{16}$).

\noindent \paragraph{Period Recovery and Factorization.}

As before, the final stage uses the classical continued fraction algorithm on the observed phase, $\hat{\phi}=7/16$. This gives the most likely denominator $r=16$, which is confirmed to be the true period of 12 modulo 221. The factorization is then completed by computing the greatest common divisor (gcd):
\begin{equation}
\label{eq:factor_221}
\text{gcd}\left(a^{r/2} \pm 1, N\right) = \text{gcd}\left(12^{16/2} \pm 1, 221\right) = \text{gcd}\left(12^{8} \pm 1, 221\right) \Rightarrow \{13, 17\}. \nonumber
\end{equation}
The algorithm successfully returns the non-trivial factors 13 and 17.

\section{Numerical Examples: Cases with Non-Standard  Initial Target State  $\ket{\psi_0}$}
In this section, we consider numerical examples with non-standard  initial target states of the form  $\ket{\psi_0}$ as described in \Cref{eq_psi_with_one} in \Cref{remark_target}, with $L \geq 2$.

\subsection{Numerical Example: Factoring $N=161$}

In this numerical example we consider factoring the integer $N=161$ with the base $a=3$.
The algorithm was configured with the following parameters: The required target qubits ($n_{\text{target}}$) is 8 ($\lceil \log_2(161) \rceil$), blocks allocation $\mathbf{m} = [3, 4, 4, 4, 4, 4, 4, 4]$, overlaps $\mathcal{T} = [0, 2, 2, 2, 2, 2, 2, 2]$ and \texttt{num\_selected\_candidates}  $=8$. The initial target state $\ket{\psi_0}$ was created  as a uniform superposition of  $L=3$ computational basis states (ref.~\Cref{eq_psi_general}). This means 
the initial target state for each block was set $\ket{\psi_0} =  \sum_{j=0}^{2} w_j \ket{a^{j} \bmod N}$, with equal weights $w_j = \frac{1}{\sqrt{3}}$ for $j=0$ to $2$. The output from each block provided candidate bitstrings and their empirical probabilities, which are shown in \Cref{tab:qpe_blocks_161}.

\begin{table}[H]
\centering
\caption{Most frequent candidates measured during the run for $N=161$ with $a=3$.}
\label{tab:qpe_blocks_161}
\renewcommand{\arraystretch}{1.35}
\begin{tabular}{c p{14cm}}
\toprule
{Block} & {Top Candidates (bitstring: empirical probability)} \\ 
\midrule
1 & \texttt{000} (33.5\%), \texttt{111} (22.8\%), \texttt{001} (22.7\%), \texttt{010} (6.3\%), \texttt{110} (6.2\%), \texttt{100} (4.2\%), \texttt{101} (2.2\%), \texttt{011} (2.2\%) \\ 
2 & \texttt{0000} (10.2\%), \texttt{0001} (9.8\%), \texttt{1111} (9.8\%), \texttt{0010} (9.0\%), \texttt{1110} (9.0\%), \texttt{1101} (7.8\%), \texttt{0011} (7.7\%), \texttt{1100} (6.3\%) \\ 
3 & \texttt{1000} (8.4\%), \texttt{0000} (8.4\%), \texttt{0110} (8.4\%), \texttt{1100} (8.3\%), \texttt{0100} (8.3\%), \texttt{0010} (8.3\%), \texttt{1110} (8.3\%), \texttt{1010} (8.3\%) \\ 
4 & \texttt{1000} (9.9\%), \texttt{0000} (9.9\%), \texttt{0111} (8.9\%), \texttt{0001} (8.9\%), \texttt{1111} (8.8\%), \texttt{1001} (8.7\%), \texttt{0110} (6.2\%), \texttt{0010} (6.2\%) \\ 
5 & \texttt{1010} (6.4\%), \texttt{0100} (6.3\%), \texttt{0000} (6.3\%), \texttt{0010} (6.3\%), \texttt{1000} (6.3\%), \texttt{0111} (6.3\%), \texttt{0001} (6.3\%), \texttt{1110} (6.2\%) \\ 
6 & \texttt{1100} (9.5\%), \texttt{0000} (9.4\%), \texttt{0100} (9.3\%), \texttt{1000} (9.3\%), \texttt{0001} (6.3\%), \texttt{0111} (6.2\%), \texttt{0011} (6.2\%), \texttt{1011} (6.2\%) \\ 
7 & \texttt{0000} (10.1\%), \texttt{1111} (9.8\%), \texttt{0001} (9.7\%), \texttt{0010} (9.1\%), \texttt{1110} (8.9\%), \texttt{0011} (7.8\%), \texttt{1101} (7.7\%), \texttt{1100} (6.3\%) \\ 
8 & \texttt{1000} (8.5\%), \texttt{0100} (8.4\%), \texttt{0010} (8.4\%), \texttt{1010} (8.3\%), \texttt{1100} (8.3\%), \texttt{0000} (8.3\%), \texttt{0110} (8.2\%), \texttt{1110} (8.2\%) \\ 
\bottomrule
\end{tabular}
\end{table}

The collected bitstrings from all QPE blocks were successfully combined using a carry-aware stitching process, yielding a full-length 17-bit candidate, \texttt{00011011000111010}. 
1111010
This bitstring corresponds to the decimal value $13{,}882$, which provides the fractional phase $\phi = \frac{13{,}882}{2^{17}} = \frac{6{,}941}{65{,}536}$. Applying the classical continued fraction algorithm to this phase successfully recovered the period $r=66$. Using this period, the final classical step involved computing $\gcd(a^{r/2} \pm 1, N) = \gcd(3^{33} \pm 1, 161)$, which successfully yielded the non-trivial factor $23$ of $N=161$.

\subsection{Numerical Example: Factoring $N = 295{,}927$}

To demonstrate the behavior and performance of the modular, windowed quantum phase estimation procedure, we consider the integer $N = 295,927$, and select the base $a = 3$, $\mathbf{m} = [4, 4, 4, 4, 4, 4, 4, 4, 4, 4, 4, 4, 4]$ and $\mathcal{T} = [0, 2, 2, 2, 2, 2, 2, 2, 2, 2, 2, 2, 2]$.
Further, for this example, \texttt{num\_selected\_candidates} was set to $=8$ and  the initial target state for each block was set using the uniform weights $w_j = \frac{1}{\sqrt{L}}$ with $L=500$ (ref.~\Cref{eq_psi_with_one}).
The output from each block with the candidate bitstrings and their empirical probabilities, which are shown in \Cref{tab:block_outcomes_295627}.

\begin{table}[H]
\centering
\caption{Most frequent candidates measured during the run for $N=295{,}627$ with $a=3$.}
\label{tab:block_outcomes_295627}
\renewcommand{\arraystretch}{1.35} 
\begin{tabular}{c p{14cm}}
\toprule
{Block} & {Top Candidates (bitstring: empirical probability)} \\
\midrule
1 & \texttt{0000} (99.0\%), \texttt{1111} (0.3\%), \texttt{0001} (0.3\%), \texttt{0010} (0.1\%), \texttt{1110} (0.1\%), \texttt{0011} (0.0\%), \texttt{1101} (0.0\%), \texttt{1100} (0.0\%) \\
2 & \texttt{0000} (95.8\%), \texttt{0001} (1.3\%), \texttt{1111} (1.3\%), \texttt{0010} (0.3\%), \texttt{1110} (0.3\%), \texttt{0011} (0.2\%), \texttt{1101} (0.2\%), \texttt{1100} (0.1\%) \\
3 & \texttt{0000} (83.0\%), \texttt{0001} (5.2\%), \texttt{1111} (5.2\%), \texttt{0010} (1.4\%), \texttt{1110} (1.4\%), \texttt{1101} (0.7\%), \texttt{0011} (0.6\%), \texttt{0100} (0.4\%) \\
4 & \texttt{0000} (41.0\%), \texttt{0001} (20.8\%), \texttt{1111} (20.7\%), \texttt{0010} (2.8\%), \texttt{1110} (2.8\%), \texttt{0011} (2.5\%), \texttt{1101} (2.5\%), \texttt{0101} (1.1\%) \\
5 & \texttt{0000} (27.2\%), \texttt{1110} (19.4\%), \texttt{0010} (19.4\%), \texttt{0100} (7.4\%), \texttt{1100} (7.3\%), \texttt{0101} (3.2\%), \texttt{1011} (3.2\%), \texttt{0011} (2.7\%) \\
6 & \texttt{0000} (27.9\%), \texttt{0111} (15.4\%), \texttt{1001} (15.4\%), \texttt{1000} (7.4\%), \texttt{0001} (7.0\%), \texttt{1111} (7.0\%), \texttt{0110} (3.1\%), \texttt{1010} (3.1\%) \\
7 & \texttt{0000} (29.6\%), \texttt{1101} (16.1\%), \texttt{0011} (16.0\%), \texttt{1011} (7.4\%), \texttt{0101} (7.3\%), \texttt{1000} (4.5\%), \texttt{1110} (4.1\%), \texttt{0010} (4.1\%) \\
8 & \texttt{0000} (32.1\%), \texttt{0101} (22.3\%), \texttt{1011} (22.3\%), \texttt{1010} (6.4\%), \texttt{0110} (6.4\%), \texttt{1100} (1.7\%), \texttt{0100} (1.7\%), \texttt{1001} (1.3\%) \\
9 & \texttt{0000} (28.7\%), \texttt{1011} (21.1\%), \texttt{0101} (21.1\%), \texttt{0110} (8.0\%), \texttt{1010} (7.9\%), \texttt{1111} (2.0\%), \texttt{0001} (2.0\%), \texttt{0100} (1.6\%) \\
10 & \texttt{0000} (27.5\%), \texttt{1011} (19.4\%), \texttt{0101} (19.2\%), \texttt{0110} (6.2\%), \texttt{1010} (6.1\%), \texttt{1000} (3.3\%), \texttt{0111} (3.2\%), \texttt{1001} (3.2\%) \\
11 & \texttt{0000} (29.2\%), \texttt{0011} (20.6\%), \texttt{1101} (20.4\%), \texttt{1010} (7.9\%), \texttt{0110} (7.8\%), \texttt{1009} (2.9\%), \texttt{0111} (2.9\%), \texttt{1100} (1.4\%) \\
12 & \texttt{0000} (27.4\%), \texttt{0011} (14.7\%), \texttt{1101} (14.6\%), \texttt{1001} (7.3\%), \texttt{0111} (7.3\%), \texttt{1100} (5.1\%), \texttt{0100} (5.1\%), \texttt{1010} (3.8\%) \\
13 & \texttt{0000} (29.4\%), \texttt{0011} (15.7\%), \texttt{1101} (15.6\%), \texttt{1011} (7.9\%), \texttt{0101} (7.8\%), \texttt{1000} (4.6\%), \texttt{1110} (4.4\%), \texttt{0010} (4.3\%) \\
\bottomrule
\end{tabular}
\end{table}

The final phase of of our Algorithm involved the stitching and classical post-processing of the collected QPE results. The successful candidate was the string $\texttt{0011000001000011111001111110}$, which corresponds to the integer value $50{,}609{,}790$. 
This yielded the fractional phase $\phi = \frac{50{,}609{,}790}{2^{28}} = \frac{25{,}304{,}895}{134{,}217{,}728}$. Applying the classical continued fraction algorithm to this phase recovered a multiple of the period as $5{,}670$ (where the true period $r$ is $1{,}890$). Using this value in the final classical step, a non-trivial factor ($541$) of $N=295{,}927 = 541 \times 547$ was successfully found.

\subsection{Numerical Example: Factoring $1{,}461{,}617$}

\begin{table}[H]
\centering
\caption{Most frequent candidates measured during the run for $N=1{,}461{,}617$ with $a=8$.}
\label{tab:block_outcomes_1461617}
\renewcommand{\arraystretch}{1.35} 
\begin{tabular}{c p{14cm}}
\toprule
{Block} & {Top Candidates (bitstring: empirical probability)} \\
\midrule
1  & \texttt{0000} (94.7\%), \texttt{1111} (1.7\%), \texttt{0001} (1.6\%), \texttt{1110} (0.4\%), \texttt{0010} (0.4\%), \texttt{1101} (0.2\%), \texttt{0011} (0.2\%), \texttt{1100} (0.1\%) \\
2  & \texttt{0000} (78.8\%), \texttt{1111} (6.6\%), \texttt{0001} (6.5\%), \texttt{1110} (1.7\%), \texttt{0010} (1.6\%), \texttt{1101} (0.8\%), \texttt{0011} (0.8\%), \texttt{1100} (0.5\%) \\
3  & \texttt{0000} (34.2\%), \texttt{0001} (21.8\%), \texttt{1111} (21.7\%), \texttt{0010} (4.9\%), \texttt{1110} (4.9\%), \texttt{1100} (1.9\%), \texttt{0100} (1.8\%), \texttt{1101} (1.7\%) \\
4  & \texttt{0000} (10.6\%), \texttt{0001} (10.2\%), \texttt{1111} (10.2\%), \texttt{1110} (9.2\%), \texttt{0010} (9.2\%), \texttt{0011} (7.9\%), \texttt{1101} (7.8\%), \texttt{1100} (6.2\%) \\
5  & \texttt{0000} (6.8\%), \texttt{1111} (6.8\%), \texttt{0001} (6.7\%), \texttt{0010} (6.7\%), \texttt{1110} (6.6\%), \texttt{0011} (6.5\%), \texttt{1101} (6.5\%), \texttt{0100} (6.3\%) \\
6  & \texttt{1110} (6.4\%), \texttt{1111} (6.4\%), \texttt{0011} (6.4\%), \texttt{0010} (6.3\%), \texttt{0001} (6.3\%), \texttt{0000} (6.3\%), \texttt{1101} (6.3\%), \texttt{1100} (6.3\%) \\
7  & \texttt{0000} (7.6\%), \texttt{0101} (7.3\%), \texttt{1011} (7.3\%), \texttt{1010} (7.0\%), \texttt{0110} (7.0\%), \texttt{0001} (6.8\%), \texttt{1111} (6.7\%), \texttt{1100} (6.3\%) \\
8  & \texttt{1110} (6.4\%), \texttt{0000} (6.4\%), \texttt{1111} (6.4\%), \texttt{1101} (6.4\%), \texttt{0001} (6.3\%), \texttt{0010} (6.3\%), \texttt{0011} (6.3\%), \texttt{0101} (6.3\%) \\
9  & \texttt{0000} (10.4\%), \texttt{0100} (10.4\%), \texttt{1100} (10.4\%), \texttt{1000} (10.0\%), \texttt{0001} (6.5\%), \texttt{0101} (6.4\%), \texttt{1111} (6.3\%), \texttt{1011} (6.3\%) \\
10 & \texttt{0000} (11.2\%), \texttt{1111} (11.0\%), \texttt{0001} (11.0\%), \texttt{1110} (9.9\%), \texttt{0010} (9.9\%), \texttt{1101} (8.2\%), \texttt{0011} (8.2\%), \texttt{0100} (6.3\%) \\
11 & \texttt{1111} (6.5\%), \texttt{0000} (6.4\%), \texttt{0001} (6.3\%), \texttt{1110} (6.3\%), \texttt{0010} (6.3\%), \texttt{1100} (6.2\%), \texttt{1101} (6.2\%), \texttt{0011} (6.2\%) \\
12 & \texttt{1111} (6.5\%), \texttt{0001} (6.4\%), \texttt{0000} (6.3\%), \texttt{0010} (6.3\%), \texttt{0100} (6.3\%), \texttt{0011} (6.3\%), \texttt{1100} (6.2\%), \texttt{1110} (6.2\%) \\
13 & \texttt{1110} (6.4\%), \texttt{0001} (6.4\%), \texttt{0000} (6.4\%), \texttt{1111} (6.3\%), \texttt{1100} (6.3\%), \texttt{0011} (6.3\%), \texttt{1011} (6.3\%), \texttt{1101} (6.2\%) \\
14 & \texttt{1111} (6.4\%), \texttt{0010} (6.4\%), \texttt{0001} (6.4\%), \texttt{1011} (6.3\%), \texttt{0000} (6.3\%), \texttt{0100} (6.3\%), \texttt{1100} (6.3\%), \texttt{0011} (6.3\%) \\
15 & \texttt{0000} (9.6\%), \texttt{1101} (9.3\%), \texttt{0011} (9.2\%), \texttt{0110} (8.4\%), \texttt{1010} (8.4\%), \texttt{0111} (7.4\%), \texttt{1001} (7.4\%), \texttt{1100} (6.4\%) \\
16 & \texttt{0000} (7.9\%), \texttt{0011} (7.6\%), \texttt{1101} (7.5\%), \texttt{1010} (7.1\%), \texttt{0110} (7.0\%), \texttt{1001} (6.6\%), \texttt{0111} (6.6\%), \texttt{0100} (6.3\%) \\
\bottomrule
\end{tabular}
\end{table}

Next,  we consider $N = 1{,}461{,}617$. We select the base $a = 8$, and use a block configuration of $\mathbf{m} = [4]^{16}$ and $\mathcal{T} = [0] +  [2]^{15}$. With 16 blocks of size $4$ and an overlap $2$, the total number of control qubits  is $4 + 15 \times (4-2) = 34$. This corresponds to a total phase space of $2^{34}$ values.

For this example, the number of top candidates collected from each block, $\texttt{num\_selected\_candidates}$, was set to $8$. The initial target state for each block was set using the uniform weights $w_j = \frac{1}{\sqrt{L}}$ with  $L=500$ ((ref.~\Cref{eq_psi_with_one})). The measured candidate bitstrings and their empirical probabilities for each of the 16 blocks are shown in \Cref{tab:block_outcomes_1461617}.

The final classical stage of the  for factoring $N=1{,}461{,}617$ involved the stitching and classical post-processing of the QPE results. After integrating the partial candidates, the successful combination  yielded the successful stitched bitstring $\texttt{0011111111001100010000110011110011}$, which corresponds to the integer value $4{,}281{,}404{,}659$. 
This resulted in the fractional phase:
$$\phi = \frac{4{,}281{,}404{,}659}{2^{34}} = \frac{4{,}281{,}404{,}659}{17{,}179{,}869{,}184}$$

Applying the classical continued fraction algorithm to this phase successfully recovered the period $r = 3{,}800$. Using this value in the final classical step $\gcd(a^{r/2} \pm 1, N) = \gcd(8^{1900} \pm 1, 1461617)$, a non-trivial factor $1201$ was successfully found. We note that in this case $N = 1201 \times 1217$.

\subsection{Numerical Example: Factoring $N=5{,}558{,}039$}

For our next example,  we consider $N = 5{,}558{,}039$. We select the base $a = 64$, and use a block configuration of $\mathbf{m} = [3] + [4]^{22}$ and $\mathcal{T} = [0] +  [2]^{22}$. With $1$ block of size $3$, followed by $22$ blocks of size $4$, and an overlap $2$, the total number of control qubits  is $3 + 22 \times (4-2) = 47$. This corresponds to a total phase space of $2^{47}$ values. 
The number of top candidates collected from each block, $\texttt{num\_selected\_candidates}$, was set to $6$ and the initial target state for each block was $\ket{\psi_0} =  \ket{\widetilde{u}_1} = \frac{1}{\sqrt{r_1}} \sum_{s=0}^{r_1-1} e^{-2\pi i \frac{s}{r_1}} \ket{a^s \bmod N}$. where $r_1 = 2^{16}$ (ref.~\Cref{eq_psi_fourier_basis}).

The candidate bitstrings and their empirical probabilities for each of the 23 blocks are shown in \Cref{tab:block_outcomes_combined_5558039}.

\renewcommand{\arraystretch}{1.35}

\begin{longtable}{c p{14cm}}
\caption{Most frequent candidates measured during the run for $N=5{,}558{,}039$ with $a=64$.}
\label{tab:block_outcomes_combined_5558039} \\

\toprule
{Block} & {Top Candidates (bitstring: empirical probability)} \\
\midrule
\endfirsthead

\multicolumn{2}{c}%
{{\bfseries Table \thetable\ (continued)}} \\
\toprule
{Block} & {Top Candidates (bitstring: empirical probability)} \\
\midrule
\endhead

\midrule
\multicolumn{2}{r}{{Continued on next page}} \\
\midrule
\endfoot

\bottomrule
\endlastfoot
1 & \texttt{000} (100.0\%), \texttt{100} (0.0\%), \texttt{111} (0.0\%), \texttt{001} (0.0\%), \texttt{010} (0.0\%), \texttt{110} (0.0\%) \\ 
2 & \texttt{0000} (100.0\%), \texttt{0001} (0.0\%), \texttt{1111} (0.0\%), \texttt{1110} (0.0\%), \texttt{1000} (0.0\%), \texttt{1101} (0.0\%) \\ 
3 & \texttt{0000} (99.9\%), \texttt{0001} (0.0\%), \texttt{1111} (0.0\%), \texttt{0010} (0.0\%), \texttt{1101} (0.0\%), \texttt{0011} (0.0\%) \\ 
4 & \texttt{0000} (99.7\%), \texttt{0001} (0.1\%), \texttt{1111} (0.1\%), \texttt{1110} (0.0\%), \texttt{0010} (0.0\%), \texttt{1101} (0.0\%) \\ 
5 & \texttt{0000} (98.7\%), \texttt{0001} (0.5\%), \texttt{1111} (0.4\%), \texttt{0010} (0.1\%), \texttt{1110} (0.1\%), \texttt{0011} (0.1\%) \\ 
6 & \texttt{0000} (90.0\%), \texttt{0001} (4.2\%), \texttt{1111} (2.2\%), \texttt{0010} (0.9\%), \texttt{1110} (0.7\%), \texttt{0011} (0.4\%) \\ 
7 & \texttt{0001} (50.4\%), \texttt{0000} (27.9\%), \texttt{0010} (6.3\%), \texttt{1111} (4.6\%), \texttt{0011} (2.2\%), \texttt{1110} (2.1\%) \\ 
8 & \texttt{0010} (71.9\%), \texttt{0011} (9.8\%), \texttt{0001} (3.3\%), \texttt{0100} (3.1\%), \texttt{0101} (2.5\%), \texttt{0111} (2.2\%) \\ 
9 & \texttt{1001} (89.8\%), \texttt{0010} (3.3\%), \texttt{1011} (1.6\%), \texttt{0100} (0.8\%), \texttt{0101} (0.6\%), \texttt{1110} (0.6\%) \\ 
10 & \texttt{0100} (72.7\%), \texttt{0101} (9.4\%), \texttt{0011} (3.6\%), \texttt{1000} (2.0\%), \texttt{1001} (1.9\%), \texttt{0110} (1.9\%) \\ 
11 & \texttt{0001} (90.5\%), \texttt{0010} (3.6\%), \texttt{0011} (1.7\%), \texttt{0100} (0.9\%), \texttt{1110} (0.6\%), \texttt{1101} (0.5\%) \\ 
12 & \texttt{0100} (83.5\%), \texttt{0101} (3.7\%), \texttt{1000} (2.9\%), \texttt{0011} (2.0\%), \texttt{1100} (1.2\%), \texttt{0110} (0.9\%) \\ 
13 & \texttt{0001} (64.0\%), \texttt{0000} (16.6\%), \texttt{0010} (6.3\%), \texttt{1111} (3.4\%), \texttt{0011} (2.3\%), \texttt{1110} (1.8\%) \\ 
14 & \texttt{0011} (59.3\%), \texttt{0010} (18.2\%), \texttt{0100} (4.3\%), \texttt{0101} (4.2\%), \texttt{0001} (2.9\%), \texttt{1000} (2.5\%) \\ 
15 & \texttt{1011} (48.5\%), \texttt{1010} (27.9\%), \texttt{1100} (4.6\%), \texttt{0101} (4.1\%), \texttt{1001} (3.8\%), \texttt{0000} (2.0\%) \\ 
16 & \texttt{1010} (69.9\%), \texttt{1011} (11.1\%), \texttt{1001} (4.3\%), \texttt{1100} (2.2\%), \texttt{0101} (2.1\%), \texttt{1111} (1.9\%) \\ 
17 & \texttt{1001} (86.9\%), \texttt{0010} (3.0\%), \texttt{1010} (1.8\%), \texttt{1011} (1.6\%), \texttt{1000} (1.1\%), \texttt{0101} (0.8\%) \\ 
18 & \texttt{0100} (39.7\%), \texttt{0101} (34.6\%), \texttt{0011} (4.7\%), \texttt{0110} (4.4\%), \texttt{1001} (4.0\%), \texttt{0010} (2.6\%) \\ 
19 & \texttt{0010} (89.5\%), \texttt{0100} (3.4\%), \texttt{0110} (1.5\%), \texttt{1000} (0.9\%), \texttt{0001} (0.7\%), \texttt{1010} (0.7\%) \\ 
20 & \texttt{1000} (68.3\%), \texttt{0111} (13.5\%), \texttt{1001} (4.2\%), \texttt{1111} (2.9\%), \texttt{0110} (2.3\%), \texttt{0000} (1.4\%) \\ 
21 & \texttt{1111} (82.4\%), \texttt{1110} (6.2\%), \texttt{1101} (2.1\%), \texttt{0000} (2.0\%), \texttt{1100} (1.3\%), \texttt{0001} (1.0\%) \\ 
22 & \texttt{1011} (67.3\%), \texttt{1100} (12.5\%), \texttt{1010} (3.8\%), \texttt{1101} (3.0\%), \texttt{0111} (2.5\%), \texttt{0010} (1.8\%) \\ 
23 & \texttt{1101} (80.3\%), \texttt{1110} (5.1\%), \texttt{1100} (2.4\%), \texttt{1010} (2.4\%), \texttt{1011} (1.8\%), \texttt{1000} (1.4\%) \\ 
\end{longtable}

Finally, the stitching and classical post-processing of the QPE results were carried out and  after integrating the partial candidates, the successful combination  resulted in the correctly stitched bitstring $\texttt{00000000000000010010001000010101001001011101100}$, corresponding to the integer value $2{,}433{,}389{,}292$.
This corresponds to the fractional phase of
$$\phi = \frac{608{,}347{,}323}{35{,}184{,}372{,}088{,}832}$$

Applying the classical continued fraction algorithm to this phase successfully recovered the period $r = 57{,}836$. Using this value in the final classical step $\gcd(a^{r/2} \pm 1, N) = \gcd(8^{57836} \pm 1, 5558039)$, a non-trivial factor $4567$ was successfully obtained. In this case, $N = 1217 \times 4567$. We also obtained the factors of $N$ successfully for other choices of $a$, for example $a=2$ where the corresponding period was found to be $r=115{,}672$.

\begin{remark}
    The correctly stitched bitstring above corresponds to the following block candidates:
    [\texttt{000, 0000, 0000, 0000, 0000, 0000, 0001, 0010, 1001, 0100, 0001, 0100, 0001, 0011, 1011, 1010, 1001, 0101, 0010, 1000, 1111, 1011, 1100}].
    The first string \texttt{000} originates from the first block, and the next six blocks are all zeros; nonzero values appear only in the remaining $17$ blocks. 
    
    This observation suggests an interesting optimization strategy. When $r$ is large, the early bits corresponding to the phase $1/r$ are likely to be zero, causing many of the initial blocks to yield zero values. By suitably tuning the weights in the Fourier-basis initial state (see~\Cref{eq_psi_fourier_basis}), this phase component can be made dominant. Consequently, one could begin analysis from the rightmost block and progressively move leftward, checking candidate bitstrings at intermediate stages and halting once the correct factor is identified (assuming that the left blocks continue to yield zeros). Such an approach could further reduce both quantum resource requirements and overall execution time of the algorithm.
\end{remark}

\subsection{Numerical Example: Factoring $N=15{,}001{,}999$}

Next,  we consider $N = 15{,}001{,}999$. We pick the base $a = 13$, and use a block configuration of $\mathbf{m} = [3] + [4]^{23}$ and $\mathcal{T} = [0] +  [2]^{23}$. With $1$ block of size $3$, followed by $23$ blocks of size $4$, and an overlap $2$. The total number of control qubits  is $3 + 23 \times (4-2) = 49$. This corresponds to a total phase space of size $2^{49}$. 
The number of top candidates picked from each block, $\texttt{num\_selected\_candidates}$, was set to $8$. The initial target state for each block was set to  $\ket{\psi_0} =  \frac{1}{\sqrt{r_j}} \sum_{s=0}^{r_j-1} e^{-2\pi i \frac{j s}{r_j}} \ket{a^s \bmod N}$,  for $j=13$ with $r_{13} = 2^{18}$ (ref.~\Cref{eq_psi_fourier_basis}).

The candidate bitstrings and their empirical probabilities for each of the $24$ blocks are shown in \Cref{tab:block_outcomes_combined_15001999}.

\begin{longtable}{c p{14cm}}
\caption{Most frequent candidates measured during the run for $N=15{,}001{,}999$ with $a=13$.}
\label{tab:block_outcomes_combined_15001999} \\

\toprule
{Block} & {Top Candidates (bitstring: empirical probability)} \\
\midrule
\endfirsthead

\multicolumn{2}{c}%
{{\bfseries Table \thetable\ (continued)}} \\
\toprule
{Block} & {Top Candidates (bitstring: empirical probability)} \\
\midrule
\endhead

\midrule
\multicolumn{2}{r}{{Continued on next page}} \\
\midrule
\endfoot
\bottomrule
\endlastfoot

1 & \texttt{000} (100.0\%), \texttt{001} (0.0\%), \texttt{111} (0.0\%), \texttt{101} (0.0\%), \texttt{010} (0.0\%), \texttt{110} (0.0\%), \texttt{100} (0.0\%) \\ 
2 & \texttt{0000} (100.0\%), \texttt{0001} (0.0\%), \texttt{1111} (0.0\%), \texttt{1110} (0.0\%), \texttt{0010} (0.0\%), \texttt{1000} (0.0\%), \texttt{1001} (0.0\%), \texttt{0111} (0.0\%) \\ 
3 & \texttt{0000} (99.9\%), \texttt{1111} (0.0\%), \texttt{0001} (0.0\%), \texttt{0010} (0.0\%), \texttt{1110} (0.0\%), \texttt{1101} (0.0\%), \texttt{0011} (0.0\%), \texttt{0100} (0.0\%) \\ 
4 & \texttt{0000} (99.2\%), \texttt{0001} (0.3\%), \texttt{1111} (0.2\%), \texttt{0010} (0.1\%), \texttt{1110} (0.1\%), \texttt{1101} (0.0\%), \texttt{0011} (0.0\%), \texttt{1100} (0.0\%) \\ 
5 & \texttt{0000} (94.1\%), \texttt{0001} (2.3\%), \texttt{1111} (1.4\%), \texttt{0010} (0.5\%), \texttt{1110} (0.4\%), \texttt{0011} (0.2\%), \texttt{1101} (0.2\%), \texttt{0100} (0.1\%) \\ 
6 & \texttt{0000} (48.1\%), \texttt{0001} (33.2\%), \texttt{1111} (4.9\%), \texttt{0010} (4.3\%), \texttt{1110} (1.9\%), \texttt{0011} (1.7\%), \texttt{1101} (1.0\%), \texttt{0100} (1.0\%) \\ 
7 & \texttt{0010} (74.6\%), \texttt{0001} (11.2\%), \texttt{0011} (3.9\%), \texttt{0000} (2.4\%), \texttt{0100} (1.6\%), \texttt{1111} (1.4\%), \texttt{0101} (1.0\%), \texttt{1110} (0.8\%) \\ 
8 & \texttt{0111} (50.2\%), \texttt{1000} (14.1\%), \texttt{1001} (10.2\%), \texttt{0100} (6.1\%), \texttt{0110} (3.7\%), \texttt{0010} (3.2\%), \texttt{0011} (2.0\%), \texttt{1101} (1.7\%) \\ 
9 & \texttt{1101} (46.5\%), \texttt{1110} (17.4\%), \texttt{1111} (8.0\%), \texttt{0101} (7.3\%), \texttt{1100} (3.8\%), \texttt{0111} (2.5\%), \texttt{1011} (2.4\%), \texttt{0100} (1.8\%) \\ 
10 & \texttt{0101} (32.4\%), \texttt{0110} (29.3\%), \texttt{0011} (10.5\%), \texttt{1011} (7.7\%), \texttt{0100} (3.9\%), \texttt{0111} (3.5\%), \texttt{1101} (2.7\%), \texttt{1110} (1.9\%) \\ 
11 & \texttt{0110} (75.3\%), \texttt{1011} (12.2\%), \texttt{1100} (3.9\%), \texttt{0101} (2.2\%), \texttt{1010} (1.2\%), \texttt{0111} (1.2\%), \texttt{1101} (0.8\%), \texttt{0100} (0.6\%) \\ 
12 & \texttt{1000} (65.2\%), \texttt{1100} (7.7\%), \texttt{1110} (7.3\%), \texttt{0111} (4.3\%), \texttt{0110} (3.6\%), \texttt{1001} (2.3\%), \texttt{1010} (2.3\%), \texttt{1101} (1.2\%) \\ 
13 & \texttt{1111} (67.1\%), \texttt{0111} (9.7\%), \texttt{0000} (8.9\%), \texttt{1000} (3.3\%), \texttt{1110} (3.0\%), \texttt{0001} (1.7\%), \texttt{1101} (1.1\%), \texttt{0110} (0.9\%) \\ 
14 & \texttt{1101} (65.6\%), \texttt{1100} (12.8\%), \texttt{1110} (6.4\%), \texttt{1111} (4.9\%), \texttt{1010} (1.7\%), \texttt{1011} (1.4\%), \texttt{0001} (1.3\%), \texttt{0000} (0.9\%) \\ 
15 & \texttt{0011} (65.4\%), \texttt{0000} (9.9\%), \texttt{1010} (5.4\%), \texttt{0100} (4.6\%), \texttt{1101} (3.4\%), \texttt{1001} (2.5\%), \texttt{0010} (2.0\%), \texttt{0110} (1.8\%) \\ 
16 & \texttt{1101} (68.4\%), \texttt{0000} (9.6\%), \texttt{0110} (5.0\%), \texttt{0011} (3.3\%), \texttt{1100} (3.2\%), \texttt{0111} (2.8\%), \texttt{1010} (1.8\%), \texttt{1110} (1.6\%) \\ 
17 & \texttt{0011} (53.9\%), \texttt{0100} (11.6\%), \texttt{0000} (9.3\%), \texttt{1010} (6.5\%), \texttt{1101} (3.4\%), \texttt{0010} (3.3\%), \texttt{0101} (2.1\%), \texttt{1001} (1.9\%) \\ 
18 & \texttt{1101} (60.0\%), \texttt{1110} (7.7\%), \texttt{0111} (6.0\%), \texttt{0001} (5.3\%), \texttt{0000} (3.3\%), \texttt{1100} (3.0\%), \texttt{0011} (2.6\%), \texttt{1111} (2.1\%) \\ 
19 & \texttt{0101} (74.4\%), \texttt{0010} (7.0\%), \texttt{1010} (3.9\%), \texttt{1011} (3.7\%), \texttt{1101} (3.4\%), \texttt{0011} (2.1\%), \texttt{1100} (0.8\%), \texttt{1110} (0.7\%) \\ 
20 & \texttt{0100} (69.6\%), \texttt{1001} (8.2\%), \texttt{1010} (7.8\%), \texttt{0101} (4.7\%), \texttt{0011} (2.3\%), \texttt{1000} (1.3\%), \texttt{0110} (1.3\%), \texttt{1011} (1.1\%) \\ 
21 & \texttt{0001} (40.1\%), \texttt{0000} (22.2\%), \texttt{0101} (7.1\%), \texttt{1000} (6.5\%), \texttt{0100} (4.2\%), \texttt{0010} (3.9\%), \texttt{1111} (3.5\%), \texttt{1101} (2.2\%) \\ 
22 & \texttt{0010} (56.8\%), \texttt{0011} (17.4\%), \texttt{0001} (11.4\%), \texttt{0100} (3.4\%), \texttt{0000} (2.1\%), \texttt{1111} (1.7\%), \texttt{0101} (1.4\%), \texttt{0110} (1.1\%) \\ 
23 & \texttt{1001} (69.5\%), \texttt{1011} (5.3\%), \texttt{0101} (4.4\%), \texttt{1100} (3.2\%), \texttt{0010} (3.1\%), \texttt{1010} (2.8\%), \texttt{0100} (2.5\%), \texttt{0000} (1.9\%) \\ 
24 & \texttt{0101} (42.9\%), \texttt{0100} (19.5\%), \texttt{1110} (7.9\%), \texttt{0010} (7.0\%), \texttt{1001} (3.9\%), \texttt{0011} (3.9\%), \texttt{0110} (3.8\%), \texttt{0111} (2.6\%) \\ 

\end{longtable}

Finally,  after the stitching and classical post-processing of the QPE results and combining the partial candidates, the successful combination  resulted in the following stitched bitstring $$\texttt{0000011110111100111010101110011011110101011100111},$$ corresponding to the integer value $17{,}015{,}952{,}501{,}479$.
This corresponds to the fractional phase of
$$\phi = \frac{17{,}015{,}952{,}501{,}479}{562{,}949{,}953{,}421{,}312}.$$

Applying the classical continued fraction algorithm to this phase successfully recovered a multiple $8{,}211{,}000$ of the period $r = 71{,}400$. Using this value in the final classical step a non-trivial factor $4999$ was successfully obtained. In this case, $N = 3001 \times 4999$.


\begin{remark}
    We have also tested our proposed modular approach for integer factorization for many other values of $N$ and $a$, including some $ N > 10^6$ and found the results to be satisfactory. 
\end{remark}

\section{Deterministic Candidate-Based Success Metric}
\label{sec:deterministic-success}

In practical implementations of windowed quantum phase estimation, each block of control qubits is
measured repeatedly to obtain an empirical frequency histogram. Due to hardware and classical
post-processing constraints, only a limited number of high-probability outcomes are retained from
each block. The purpose of this section is to formulate a deterministic notion of algorithmic
success that is consistent with the stitching
procedure.

Let the target register be in the state \(\ket{\psi_0} = \sum_{k=0}^{r-1} c_k \ket{u_k}\), where
\(U \ket{u_k} = e^{2\pi i k/r} \ket{u_k}\) and \(\phi_k = k/r\). The control register is divided into
\(B\) blocks of sizes \(m_1,\ldots,m_B\), and a measurement of block \(i\) yields an outcome
\(y_i \in \{0,\ldots,2^{m_i}-1\}\). For the eigenstate \(\ket{u_k}\), the probability of obtaining
\(y_i\) in block \(i\) is
\begin{equation}
    P_{m_i}(y_i; k, r)
=
\frac{1}{2^{2m_i}}
\left|
\frac{
\sin \bigl(\pi 2^{m_i}(\frac{k}{r} - \frac{y_i}{2^{m_i}})\bigr)
}{
\sin \bigl(\pi (\frac{k}{r} - \frac{y_i}{2^{m_i}})\bigr)
}
\right|^2 .
\end{equation}
For the superposition \(\ket{\psi_0}\), the induced blockwise distribution is
\begin{equation}
    P_{m_i}(y_i; \psi_0) = \sum_{k=0}^{r-1} |c_k|^2\, P_{m_i}(y_i; k, r),
\end{equation}
 which determines the
experimental frequency histogram.

From this histogram, one selects the most frequently observed outcomes. We denote the ordered list
for block \(i\) by \(\mathcal{S}_i = (y_i^{(1)}, y_i^{(2)}, \ldots, y_i^{(N_i^{\mathrm{sel}})})\),
where \(y_i^{(1)}\) has the highest empirical frequency. In \Cref{alg:main}, the corresponding
quantity appears as \texttt{num\_selected\_candidates} and is taken to be identical for all blocks for simplicity.
The notation \(N_i^{\mathrm{sel}}\) used here accommodates block-dependent values. The case used in
\Cref{alg:main}  corresponds to \(N_i^{\mathrm{sel}} =
\texttt{num\_selected\_candidates}\) for $i=1,2, \ldots   B$.


A global stitched candidate is represented as  \(\hat{y} := g(y_1,\ldots,y_B)\) with  
\(y_i \in \mathcal{S}_i\), and the function $g$ captures the stitching procedure described in  
\Cref{alg:stitch}. After stitching and post-processing, the retained set of sequences is
\(\mathcal{C}_{\max} := \{\hat{y}_1, \hat{y}_2, \ldots, \hat{y}_{\mathrm{max\_limit\_combos}}\}\). For
each eigenphase \(\phi_k = k/r\), the ideal stitched sequence is
\(\hat{y}^*(k,r) = (\hat{y}_1^*,\ldots,\hat{y}_B^*)\), where \(\hat{y}_i^*\) is the best \(m_i\)-bit approximation of
\(2^\kappa \phi_k\) on block \(i\) in the sense that \(|\,\hat{y}_i^*/2^{m_i} - 2^\kappa \phi_k\,|\) is
minimized.

We define the deterministic success count
\[
P_{\mathrm{success}}
=
\sum_{k=0}^{r-1}
\mathbf{1}\!\left\{\hat{y}^*(k,r) \in \mathcal{C}_{\mathrm{max}}\right\},
\]
which satisfies \(P_{\mathrm{success}} \in \{0,\ldots,r\}\). The procedure is regarded as successful
whenever \(P_{\mathrm{success}} \ge 1\), meaning that at least one eigenphase has its ideal
blockwise sequence represented in the stitched candidate set.

Once the candidate set \(\mathcal{C}_{\mathrm{max}}\) is fixed, the quantity
\(P_{\mathrm{success}}\) is entirely deterministic and independent of the amplitudes \(|c_k|^2\).
Probabilistic considerations enter only through the sampling noise that determines the content of
the lists \(\mathcal{S}_i\). The overlap parameter in the stitching algorithm enforces compatibility across adjacent blocks by
comparing the shared bit regions of successive outcomes. When the empirical histograms are well
resolved, this consistency requirement ensures that only sequences aligned with the true eigenphase
survive the stitching step, thereby suppressing spurious candidates. Further, if one eigenstate carries sufficiently large weight (for example,
\(|c_k|^2 > 1/N_i^{\mathrm{sel}}\)) and the empirical histograms are well resolved, the correct
blockwise outcomes will be included in the selected candidate sets with near certainty.

\section{Computational Complexity}
\label{sec:complexity}

Asymptotically, the quantum circuit depth arising from our proposed modular approach for integer factorization remains the same as that in standard Shor's algorithm. However, the quantum circuits for each block are shallower in comparison to the quantum circuit corresponding to the standard implementation of Shor's algorithm. In our proposed modular framework, each circuit block has depth has depth 
$\mathcal{O}(m_{\max}\cdot \mathrm{Depth}(U))$. We further note that, in our proposed framework, the qubit requirement for the work register is the same as that of the standard Shor’s algorithm.

We also note that the preparation of the non-standard initial target state $\ket{\psi_0}$, as defined in \Cref{eq_psi_general}, typically incurs a cost depending on $L$. If $L$ is chosen to be a small fixed number (e.g., $L=4$ or $L=500$), and a uniform superposition of states is prepared (i.e., $w_j = 1/\sqrt{L}$), then this state preparation cost may be very small compared to the cost of the controlled modular exponentiation operations involved in the QPE blocks.

A key advantage of our proposed approach is a significant reduction in the number of phase (or counting) qubits per circuit block.
For an $n$-bit integer, the size of the phase register in the standard algorithm is approximately $2n+1$, whereas in the modular framework it is reduced to $m_{\max}$, where $m_{\max} = \max(m_1, m_2, \ldots, m_B)$ and $m_k$ is the number of qubits allocated to the $k$-th block. In practice, $m_{\max}$ can be chosen to be very small, for example 3 or 4 phase qubits were sufficient for the computational examples considered (ref.~\Cref{sec:results}). 
At RSA-2048 scale ($n=2048$), the phase register requirement drops from 4097 to $m_{\max}$.

The classical cost arises from carry-aware stitching of block outcomes combined 
with continued fraction expansion. 
The use of overlap and pruning thresholds 
(\texttt{max\_limit\_combos}) restrict the search to a small set, keeping the 
classical overhead negligible in practice.

\section{Conclusion}
\label{sec:conclusion}

We have introduced a modular, windowed formulation of Shor’s algorithm that restructures
phase estimation into $B$ independent, shallow blocks that can be executed sequentially or in
parallel, followed by carry-aware classical stitching (ref.~\Cref{alg:main,alg:run_block,alg:stitch,alg:recover_corrected}).
The result is a significant reduction of the counting qubit requirement (corresponding to the phase register) from approximately
$2n+1$ to $m_{\max}$, where $m_{\max} = \max (m_1, m_2, \ldots, m_B )$ and $m_k$ is the number of qubits allocated to the $k$-th block. Here, $m_{\max}$ can be chosen to be small, e.g.\ 3 or 4 qubits were found to be sufficient for the computational examples considered (ref.~\Cref{sec:results}).  Further, these independent blocks are amenable to parallelization,
which allows for a shallower effective circuit depth and significantly lower overall
execution time, making the approach far more viable for near-term quantum hardware than
the standard Shor’s algorithm. The number of qubits required for the work register in our approach remains the same as in the standard Shor's algorithm.

One of the key ingredients of this framework is the overlap mechanism. By enforcing
consistency across overlapping phase windows up to a single-bit carry, the stitching
procedure can reliably reconstruct full-length phase candidates obtained from independent blocks. This overlap-based redundancy improves robustness against noise and prunes
inconsistent candidates, while zero-overlap configurations can still succeed in some
settings with small block counts.

Our proposed modular, windowed formulation of Shor's algorithm
directly mitigates one of the principal limitations of NISQ
devices, namely the restricted number of high-quality qubits and limited coherence times.
The modular phase-estimation layer is complementary to recent advances in modular arithmetic
optimizations, and together these approaches offer cumulative improvements in the overall
resource efficiency of quantum factoring. By transforming Shor’s algorithm into a sequence
of shallow, parallelizable blocks with lightweight classical reconstruction, our proposed formulation could provide a practical approach for quantum factoring on near-term
and future fault-tolerant devices.


\end{document}